\documentstyle[12pt,axodraw]{article}
\begin{document}

\pagestyle{empty}

\newcommand{\bfig}{\begin{center}\begin{picture}}
\newcommand{\efig}[1]{\end{picture}\\{\small #1}\end{center}}
\newcommand{\flin}[2]{\ArrowLine(#1)(#2)}
\newcommand{\wlin}[2]{\DashLine(#1)(#2){2}}
\newcommand{\zlin}[2]{\DashLine(#1)(#2){5}}
\newcommand{\glin}[3]{\Photon(#1)(#2){2}{#3}}
\newcommand{\lin}[2]{\Line(#1)(#2)}
\newcommand{\sof}{\SetOffset}
\newcommand{\bmip}[2]{\begin{minipage}[t]{#1pt}\bfig(#1,#2)}
\newcommand{\emip}[1]{\efig{#1}\end{minipage}}
\newcommand{\putk}[2]{\Text(#1)[r]{$p_{#2}$}}
\newcommand{\putp}[2]{\Text(#1)[l]{$p_{#2}$}}
\newcommand{\bq}{\begin{equation}}
\newcommand{\eq}{\end{equation}}
\newcommand{\bqa}{\begin{eqnarray}}
\newcommand{\eqa}{\end{eqnarray}}
\newcommand{\nl}{\nonumber \\}
\newcommand{\eqn}[1]{Eq.(\ref{#1})}
\newcommand{\ibidem}{{\it ibidem\/},}
\newcommand{\into}{\;\;\to\;\;}
\newcommand{\epl}{e^+}
\newcommand{\emn}{e^-}
\newcommand{\nue}{\nu_e}
\newcommand{\nueb}{\bar{\nu}_e}
\newcommand{\mpl}{\mu^+}
\newcommand{\mmn}{\mu^-}
\newcommand{\num}{\nu_{\mu}}
\newcommand{\numb}{\bar{\nu}_{\mu}}
\newcommand{\tpl}{\tau^+}
\newcommand{\tmn}{\tau^-}
\newcommand{\nut}{\nu_{\tau}}
\newcommand{\nutb}{\bar{\nu}_{\tau}}
\newcommand{\ubar}{\bar{u}}
\newcommand{\dbar}{\bar{d}}
\newcommand{\cbar}{\bar{c}}
\newcommand{\sbar}{\bar{s}}
\newcommand{\bbar}{\bar{b}}
\newcommand{\ww}[2]{\langle #1 #2\rangle}
\newcommand{\wws}[2]{\langle #1 #2\rangle^{\star}}
\newcommand{\smod}{\tilde{\sigma}}
\newcommand{\dilog}[1]{\mbox{Li}_2\left(#1\right)}
\newcommand{\umu}{^{\mu}}
\newcommand{\cjg}{^{\star}}
\newcommand{\lgn}[1]{\log\left(#1\right)}
\newcommand{\si}{\sigma}
\newcommand{\sit}{\sigma_{tot}}
\newcommand{\sqs}{\sqrt{s}}
\newcommand{\sih}{\hat{\sigma}}
\newcommand{\sith}{\hat{\sigma}_{tot}}
\newcommand{\p}[1]{{\scriptstyle{\,(#1)}}}
\newcommand{\res}[3]{$#1 \pm #2~~\,10^{-#3}$}
\newcommand{\rrs}[2]{\multicolumn{1}{l|}{$~~~.#1~~10^{#2}$}}
\newcommand{\err}[1]{\multicolumn{1}{l|}{$~~~.#1$}}
\newcommand{\ru}[1]{\raisebox{-.2ex}{#1}}
\hyphenation{brems-strahl-ung}

\title{
\vspace{-4cm}
\begin{flushright}
{\large  INLO-PUB-1/94}\\
{\large  NIKHEF-H/94-08}
\end{flushright}
\vspace{4cm}
All electroweak four fermion processes in electron-positron collisions
\footnote{This research has been partly supported by EU under contract
number\\ CHRX-CT-0004.}}
\author{F.A.~Berends\thanks{email address: berends@rulgm0.LeidenUniv.nl}
{}~and R.~Pittau\thanks{email address: pittau@rulgm0.LeidenUniv.nl}\\
        Instituut-Lorentz, Leiden, The Netherlands\\
        \and
        R.~Kleiss\thanks{email address: t30@nikhefh.nikhef.nl}\\
        NIKHEF-H, Amsterdam, The Netherlands}
\maketitle
\begin{abstract}
This paper studies the electroweak production of all possible
four fermion states in $e^+ e^-$ collisions. Since the methods employed
to evaluate the complete matrix elements and phase space are
 very general, all four fermion final states in which the charged particles
 are detected can be considered. Also all kinds of experimental cuts can
be imposed. With the help of the constructed event generator a large
number of illustrative results is obtained, which show the relevance of
 backgrounds to a number of signals. For LEP 200 the W-pair signal and its
background are discussed, for higher energies also Z-pair
and single W and Z signals and backgrounds are presented.
\end{abstract}

\newpage
\pagestyle{plain}
\setcounter{page}{1}

\section{Introduction}
In electron-positron collisions at a few hundred GeV, various gauge-boson
production processes will be studied. At LEP 200, the reaction
\bq
\epl\;\emn \into W^+\;W^-        \label{wpair}
\eq
will be measured, while at somewhat higher energies also
\bqa
\epl\emn  & \into & Z\;Z\;\;, \label{zpair}               \\
\epl\emn  & \into & W\;e\;\nu_e\;\;, \label{singlew}      \\
\epl\emn  & \into & Z\;\epl\;\emn \;\;, \label{singleze}     \\
\epl\emn  & \into & Z\;\nue\;\nueb\;\;, \label{singleznu}
\eqa
will come under consideration. The double-gauge boson cross sections
decrease, but the single-gauge boson ones slowly increase,
with increasing beam energy. Around 500 GeV, the cross sections
for (\ref{wpair}), (\ref{singlew}) and (\ref{singleze}) all have about
the same magnitude, while those for (\ref{zpair}) and (\ref{singleznu})
are lower by roughly an order of magnitude \cite{hagiwara} (it should
be noted that there exists a number of rarer processes in which a single
vector boson is produced).

At LEP 200, the process (\ref{wpair}) will be used for an accurate
measurement of the $W$ mass \cite{lep200review}. Another issue is the
test of the non\--A\-be\-li\-an couplings in reaction (\ref{wpair}), and,
at higher energies, also in the other processes. The effects of
non-standard couplings between the gauge-bosons have been extensively
discussed in the literature \cite{nonstandard}. These involve terms
additional to the standard non\--A\-be\-li\-an vertices, as well as altogether
new vertices, like $ZZ\gamma$, that would also affect processes independent
of non\--A\-be\-li\-an couplings such as (\ref{zpair}) and (\ref{singleze}).

It is clear, that for these studies accurate predictions for the
cross sections (\ref{wpair})-(\ref{singleznu}) are needed. Therefore,
the effects of radiative corrections should be known. Again, a number of
such studies is available \cite{radcor4,radcor5}. For W-pair
production,
the radiative corrections have been studied most completely.
Corrections for the total cross section range from -20 to +3 per cent
when the collision energy increases from 170 to 500 GeV \cite{radcor4}.

In addition to the radiative corrections, there is another problem.
The actually measurable final states in the above reactions are
not the gauge bosons themselves, but rather their decay products. Thus,
all the above reactions are just special cases of
\bq
\epl\;\emn \into \mbox{4 fermions}\;\;. \label{fourferm}
\eq
To a specific final state, many Feynman diagrams can contribute. Some of
them are related to the reactions (\ref{wpair})-(\ref{singleznu});
others are not. To distinguish them, they will be called {\em signal\/}
and {\em background\/} diagrams, respectively. The physics issues
mentioned above necessitate a good knowledge of the relative importance
of signal and background diagrams. It may also happen, that different
four-fermion processes lead to the same {\em detectable\/}
final state as does the signal. Such a process will be called a {\em
non-interfering\/} background.\\

The aim of this paper is to investigate, at tree level,
the contribution of signal and background diagrams to the various
four-fermion final states. In its full generality, this task is
complicated for two reasons.

One reason is that, for a specified final state, the number of Feynman
diagrams can be quite large. The matrix elements then become complicated,
and we need an efficient method to compute them. In the present paper,
spinorial techniques for helicity amplitudes
\cite{spinors6}-\cite{spinors8} are used.
Such techniques are most efficient when the fermions are massless.
Since we are primarily interested in final-state configurations where
charged leptons are visible at (relatively) large angles to the beams
and between each other, collinear situations are absent. We therefore
assume all the fermions to be massless. Note that this implies the
{\em absence\/} of diagrams where a Higgs boson couples to the fermions;
but, although we cannot compute Higgs signals in our approach, we can
at least reliably estimate the background.

The second complication arises from the peaking structure.
There are many different peaks in the multidifferential cross section,
each of which finds its best description in terms of a characteristic
variable. Hence, the number of characteristic variables may be much
larger than the number of independent kinematical variables. We solve this
problem by using a multichannel Monte Carlo method \cite{multi9,multi10}.
The result of our work, then, is an event generator for four-fermion
final states.

The Monte Carlo method allows us to evaluate four-fermion cross sections
under the imposition of all kinds of cuts. In this paper, we shall do
this for a number of illustrative examples, indicating the effects
of the backgrounds on the measured signals. For realistic experimental
situations, other effects, like quark fragmentation and hadronization,
as well as detector characteristics, must be considered. Nevertheless, the
examples below already show that background diagrams can contribute
4 to 10 per cent to certain four-fermion final states.

It should be noted that the literature contains a number of papers
\cite{aeppli}-\cite{ref14} where also background effects are
calculated. None of these considers all possible final states. Moreover,
our emphasis on obtaining an efficient event generator allows for
more experimentally precise cross section estimates than the results
under necessarily simplified phase-space cuts.\\

The outline of the paper is as follows. Section 2 lists {\em all\/}
four-fermion final states for $\epl\emn $ collisions, together with the
number of Feynman diagrams. The next section describes the
matrix element computation. In section 4, the phase-space distributions
are discussed, and illustrative examples of various cross sections
are given in section 5. In the last section, we summarize our
conclusions. In an appendix toy model cross sections are given, which serve
as checks and estimates of phase space integrals.

\section{Signals and Backgrounds}
Since the $W^{\pm}$ and $Z$ decay into leptons and quarks, one may classify
the four-fermion final states into leptonic, semileptonic and hadronic
final states. The following decays for the $W^+$ and $Z$ are considered:
\bqa
W^+ & \into & \epl\nue\;,\;\mpl\num\;,\;\tpl\nut\;\;, \label{wlept} \\
W^+ & \into & u\dbar\;,\;c\sbar\;\;, \label{whadr}\\
Z & \into & \epl\emn\;,\;\mpl\mmn\;,\;\tpl\tmn\;,\;
            \nue\nueb\;,\;\num\numb\;,\;\nut\nutb\;\;, \label{zlept}\\
Z & \into & u\ubar\;,\;d\dbar\;,\;c\cbar\;,\;s\sbar\;,\;b\bbar\;\;.
    \label{zhadr}
\eqa
In principle, the $W$ knows more quark decays than (\ref{whadr}).
Those that involve a $b$ quark are extremely rare, and can safely
be neglected. The decay $u\sbar$ can, usually, experimentally not be
distinguished from $u\dbar$; and the same holds for $c\sbar$ and $c\dbar$.
The unitarity of the Cabibbo matrix therefore ensures that the use of only
(\ref{whadr}) will give the experimentally relevant predictions correctly.

\bfig(150,60)
\sof(10,10)
\flin{0,0}{20,10} \flin{20,10}{20,30} \flin{20,30}{0,40}
\wlin{20,10}{40,0} \wlin{20,30}{40,40}
\sof(70,10)
\flin{0,0}{10,20} \flin{10,20}{0,40}
\glin{10,20}{30,20}{4} \wlin{30,20}{40,0} \wlin{30,20}{40,40}
\sof(130,10)
\flin{0,0}{10,20} \flin{10,20}{0,40}
\zlin{10,20}{30,20} \wlin{30,20}{40,0} \wlin{30,20}{40,40}
\efig{Figure 1:
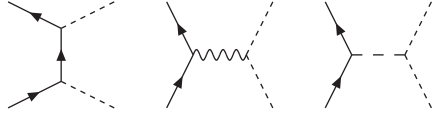
signal diagrams for $e^+e^-\to W^+W^-$. Here, and in the
following figures, solid arrowed lines stand for fermions,
dotted lines denote $W$'s, dashed lines $Z$'s, and wavy lines stand
for photons.}
The diagrams for the reactions (\ref{wpair})-(\ref{singleznu}) are
given in figs.1-5.
In these diagrams, the vector  bosons are assumed
to be stable. It is convenient to use names for certain types of
diagrams, as they usually show a characteristic peaking behaviour.
The first diagram of fig.1 is called a {\em conversion\/} diagram, and
the other ones in fig.1 are called {\em annihilation\/} diagrams.
Thus, $ZZ$ production (fig.2) consists of conversion diagrams only,
whereas $Z\epl\emn$ has two conversion diagrams.
\bfig(120,60)(0,0)
\sof(10,10)
\flin{0,0}{20,10} \flin{20,10}{20,30} \flin{20,30}{0,40}
\zlin{20,10}{40,0} \zlin{20,30}{40,40}
\sof(70,10)
\flin{0,0}{20,10} \flin{20,10}{20,30} \flin{20,30}{0,40}
\zlin{20,30}{40,0} \zlin{20,10}{40,40}
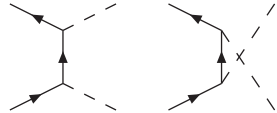
\efig{Figure 2: signal diagrams for $e^+e^-\to ZZ$.}
In single $W$
production we encounter new types of diagrams. The first and second diagram
of fig.3 is called a {\em fusion\/} diagram, the next 4 ones are called
{\em bremsstrahlung\/} diagrams. The latter represent bremsstrahlung of
(in this case) a $W$ from Bhabha-like scattering. Similar bremsstrahlung
graphs (now with a $Z$ radiated off) occur in single $Z$ production,
depicted in fig.4.
\bfig(300,120)
\sof(10,70)
\flin{0,0}{20,10} \flin{20,10}{40,0}
\flin{40,40}{20,30} \flin{20,30}{0,40}
\glin{20,10}{25,20}{2} \wlin{20,30}{25,20} \wlin{25,20}{45,20}
\sof(70,70)
\flin{0,0}{20,10} \flin{20,10}{40,0}
\flin{40,40}{20,30} \flin{20,30}{0,40}
\zlin{20,10}{20,20} \wlin{20,30}{20,20} \wlin{20,20}{45,20}
\sof(130,70)
\flin{0,0}{20,10} \flin{20,10}{40,0}
\flin{40,40}{20,30} \flin{20,30}{10,35} \flin{10,35}{0,40}
\zlin{20,10}{20,30} \wlin{10,35}{30,45}
\sof(190,70)
\flin{0,0}{20,10} \flin{20,10}{40,0}
\flin{40,40}{30,35} \flin{30,35}{20,30} \flin{20,30}{0,40}
\glin{20,10}{20,30}{4} \wlin{30,35}{50,25}
\sof(250,70)
\flin{0,0}{20,10} \flin{20,10}{40,0}
\flin{40,40}{30,35} \flin{30,35}{20,30} \flin{20,30}{0,40}
\zlin{20,10}{20,30} \wlin{30,35}{50,25}
\sof(40,10)
\flin{0,0}{20,10} \flin{20,10}{30,5} \flin{30,5}{40,0}
\flin{40,40}{20,30} \flin{20,30}{0,40}
\wlin{20,10}{20,30} \wlin{30,5}{50,15}
\sof(100,10)
\flin{0,0}{10,20} \flin{10,20}{0,40}
\flin{20,20}{30,0} \flin{25,30}{20,20} \flin{30,40}{25,30}
\glin{10,20}{20,20}{2} \wlin{25,30}{45,20}
\sof(160,10)
\flin{0,0}{10,20} \flin{10,20}{0,40}
\flin{20,20}{30,0} \flin{25,30}{20,20} \flin{30,40}{25,30}
\zlin{10,20}{20,20} \wlin{25,30}{45,20}
\sof(220,10)
\flin{0,0}{10,20} \flin{10,20}{0,40}
\flin{25,10}{30,0} \flin{20,20}{25,10} \flin{30,40}{20,20}
\zlin{10,20}{20,20} \wlin{25,10}{45,20}
\efig{Figure 3: signal diagrams for $e^+e^-\to W^+e^-\bar{\nu}_e$}

\bfig(380,200)
\sof(10,130)
\flin{0,0}{20,10} \flin{20,10}{40,0}
\flin{40,40}{20,30} \flin{20,30}{10,35} \flin{10,35}{0,40}
\zlin{10,35}{30,45} \glin{20,10}{20,30}{4}
\sof(70,130)
\flin{0,0}{20,10} \flin{20,10}{40,0}
\flin{40,40}{20,30} \flin{20,30}{10,35} \flin{10,35}{0,40}
\zlin{10,35}{30,45} \zlin{20,10}{20,30}
\sof(130,130)
\flin{0,0}{20,10} \flin{20,10}{40,0}
\flin{40,40}{30,35} \flin{30,35}{20,30} \flin{20,30}{0,40}
\zlin{30,35}{50,25} \glin{20,10}{20,30}{4}
\sof(190,130)
\flin{0,0}{20,10} \flin{20,10}{40,0}
\flin{40,40}{30,35} \flin{30,35}{20,30} \flin{20,30}{0,40}
\zlin{30,35}{50,25} \zlin{20,10}{20,30}
\sof(250,130)
\flin{0,0}{10,5} \flin{10,5}{20,10} \flin{20,10}{40,0}
\flin{40,40}{20,30} \flin{20,30}{0,40}
\zlin{10,5}{30,-5} \glin{20,10}{20,30}{4}
\sof(310,130)
\flin{0,0}{10,5} \flin{10,5}{20,10} \flin{20,10}{40,0}
\flin{40,40}{20,30} \flin{20,30}{0,40}
\zlin{10,5}{30,-5} \zlin{20,10}{20,30}
\sof(10,70)
\flin{0,0}{20,10} \flin{20,10}{30,5} \flin{30,5}{40,0}
\flin{40,40}{20,30} \flin{20,30}{0,40}
\zlin{30,5}{50,15} \glin{20,10}{20,30}{4}
\sof(70,70)
\flin{0,0}{20,10} \flin{20,10}{30,5} \flin{30,5}{40,0}
\flin{40,40}{20,30} \flin{20,30}{0,40}
\zlin{30,5}{50,15} \zlin{20,10}{20,30}
\sof(130,70)
\flin{0,0}{10,20} \flin{10,20}{0,40}
\flin{30,40}{25,30} \flin{25,30}{20,20} \flin{20,20}{30,0}
\glin{10,20}{20,20}{2} \zlin{25,30}{45,20}
\sof(190,70)
\flin{0,0}{10,20} \flin{10,20}{0,40}
\flin{30,40}{25,30} \flin{25,30}{20,20} \flin{20,20}{30,0}
\zlin{10,20}{20,20} \zlin{25,30}{45,20}
\sof(250,70)
\flin{0,0}{10,20} \flin{10,20}{0,40}
\flin{30,40}{20,20} \flin{20,20}{25,10} \flin{25,10}{30,0}
\glin{10,20}{20,20}{2} \zlin{25,10}{45,20}
\sof(310,70)
\flin{0,0}{10,20} \flin{10,20}{0,40}
\flin{30,40}{20,20} \flin{20,20}{25,10} \flin{25,10}{30,0}
\zlin{10,20}{20,20} \zlin{25,10}{45,20}
\sof(130,10)
\flin{0,0}{20,10} \flin{20,10}{20,30} \flin{20,30}{0,40}
\glin{20,30}{30,30}{2} \flin{45,15}{30,30} \flin{30,30}{45,45}
\zlin{20,10}{40,0}
\sof(190,10)
\flin{0,0}{20,10} \flin{20,10}{20,30} \flin{20,30}{0,40}
\glin{20,10}{30,10}{2} \flin{45,-5}{30,10} \flin{30,10}{45,25}
\zlin{20,30}{40,40}
\efig{Figure 4: signal diagrams for $e^+e^-\to Z e^+e^-$}
\bfig(240,120)
\sof(10,70)
\flin{0,0}{20,10} \flin{20,10}{40,0}
\flin{40,40}{20,30} \flin{20,30}{10,35} \flin{10,35}{0,40}
\zlin{10,35}{30,45} \wlin{20,10}{20,30}
\sof(70,70)
\flin{0,0}{20,10} \flin{20,10}{40,0}
\flin{40,40}{30,35} \flin{30,35}{20,30} \flin{20,30}{0,40}
\zlin{30,35}{50,25} \wlin{20,10}{20,30}
\sof(130,70)
\flin{0,0}{10,5} \flin{10,5}{20,10} \flin{20,10}{40,0}
\flin{40,40}{20,30} \flin{20,30}{0,40}
\zlin{10,5}{30,-5} \wlin{20,10}{20,30}
\sof(190,70)
\flin{0,0}{20,10} \flin{20,10}{30,5} \flin{30,5}{40,0}
\flin{40,40}{20,30} \flin{20,30}{0,40}
\zlin{30,5}{50,15} \wlin{20,10}{20,30}
\sof(40,10)
\flin{0,0}{20,10} \flin{20,10}{40,0}
\flin{40,40}{20,30} \flin{20,30}{0,40}
\wlin{20,10}{25,20} \wlin{20,30}{25,20} \zlin{25,20}{45,20}
\sof(100,10)
\flin{0,0}{10,20} \flin{10,20}{0,40}
\flin{30,40}{25,30} \flin{25,30}{20,20} \flin{20,20}{30,0}
\zlin{10,20}{20,20} \zlin{25,30}{45,20}
\sof(160,10)
\flin{0,0}{10,20} \flin{10,20}{0,40}
\flin{30,40}{20,20} \flin{20,20}{25,10} \flin{25,10}{30,0}
\zlin{10,20}{20,20} \zlin{25,10}{45,20}
\efig{Figure 5: signal diagrams for $e^+e^-\to Z\nu_e\bar{\nu}_e$}

Note that there is a number of other single $W$ or $Z$ production
processes. They are of the form
\bqa
\epl\emn & \into & \mmn\;\numb\;W^+\;\;,\label{singlewmu}\\
\epl\emn & \into & d\;\ubar\;W^+\;\;,\label{singlewhadr}\\
\epl\emn & \into & \mpl\;\mmn\;Z\;\;,\label{singlezmu}\\
\epl\emn & \into & \num\;\numb\;Z\;\;,\label{singleznum}\\
\epl\emn & \into & q\;\bar{q}\;Z\;\;.\label{singlezhadr}
\eqa
These reactions receive contributions from the annihilation graphs
of figs.3-5. Reactions (\ref{singlezmu}) and (\ref{singlezhadr}) are
also possible through the conversion diagrams of fig.4.

When the bosons decay, one gets the four-fermion final states. In figs.1-5
the decay fermions are attached to the external gauge bosons. In cases
with identical fermions, more diagrams can arise. For instance, when
both $Z$'s decay into $\epl\emn$, fig.2 will lead to 4 diagrams. Besides
the signal diagrams of figs.1-5, there are always a number of background
diagrams. The largest numbers occur for $\epl\epl\emn\emn$
and $\epl\emn\nue\nueb$, the former with only neutral-current
interactions. The background for $\epl\emn\nue\nueb$ contains
two bremsstrahlung graphs, one fusion graph, and 9 {\em multiperipheral}
diagrams (fig. 6).
\bfig(390,140)
\sof(130,85)
\flin{0,0}{20,10} \flin{20,10}{40,0}
\flin{40,40}{20,30} \flin{20,30}{0,40}
\wlin{20,10}{25,20} \wlin{20,30}{25,20} \glin{25,20}{35,20}{2}
\flin{55,30}{35,20} \flin{35,20}{55,10}
\sof(10,85)
\flin{0,0}{20,10} \flin{20,10}{40,0} \wlin{20,10}{20,20}
\flin{40,25}{20,20} \flin{20,20}{10,25} \flin{10,25}{0,30}
\glin{10,25}{20,30}{2} \flin{40,35}{20,30} \flin{20,30}{30,45}
\sof(70,85)
\flin{0,10}{10,15} \flin{10,15}{20,20} \flin{20,20}{40,15}
\flin{40,40}{20,30} \flin{20,30}{0,40} \wlin{20,20}{20,30}
\glin{10,15}{20,10}{2} \flin{30,-5}{20,10} \flin{20,10}{40,5}
\sof(190,80)
\flin{0,0}{15,10} \flin{15,10}{30,0}
\flin{30,50}{15,40} \flin{15,40}{0,50}
\flin{40,40}{25,30} \flin{25,30}{25,20} \flin{25,20}{40,10}
\wlin{15,10}{25,20} \wlin{15,40}{25,30}
\sof(250,80)
\flin{0,0}{15,10} \flin{15,10}{30,0}
\flin{30,50}{15,40} \flin{15,40}{0,50}
\flin{40,40}{25,30} \flin{25,30}{25,20} \flin{25,20}{40,10}
\zlin{15,10}{25,20} \zlin{15,40}{25,30}
\sof(310,80)
\flin{0,0}{15,10} \flin{15,10}{30,0}
\flin{30,50}{15,40} \flin{15,40}{0,50}
\flin{25,30}{40,40} \flin{25,20}{25,30} \flin{40,10}{25,20}
\zlin{15,10}{25,20} \zlin{15,40}{25,30}
\sof(10,10)
\flin{0,0}{15,10} \flin{15,10}{30,0}
\flin{30,50}{15,40} \flin{15,40}{0,50}
\flin{40,40}{25,30} \flin{25,30}{25,20} \flin{25,20}{40,10}
\wlin{15,10}{25,20} \zlin{15,40}{25,30}
\sof(75,10)
\flin{0,0}{15,10} \flin{15,10}{30,0}
\flin{30,50}{15,40} \flin{15,40}{0,50}
\flin{40,40}{25,30} \flin{25,30}{25,20} \flin{25,20}{40,10}
\zlin{15,10}{25,20} \wlin{15,40}{25,30}
\sof(140,10)
\flin{0,0}{15,10} \flin{15,10}{30,0}
\flin{30,50}{15,40} \flin{15,40}{0,50}
\flin{25,30}{40,40} \flin{25,20}{25,30} \flin{40,10}{25,20}
\wlin{15,10}{25,20} \glin{15,40}{25,30}{3}
\sof(205,10)
\flin{0,0}{15,10} \flin{15,10}{30,0}
\flin{30,50}{15,40} \flin{15,40}{0,50}
\flin{25,30}{40,40} \flin{25,20}{25,30} \flin{40,10}{25,20}
\wlin{15,10}{25,20} \zlin{15,40}{25,30}
\sof(260,10)
\flin{0,0}{15,10} \flin{15,10}{30,0}
\flin{30,50}{15,40} \flin{15,40}{0,50}
\flin{25,30}{40,40} \flin{25,20}{25,30} \flin{40,10}{25,20}
\glin{15,10}{25,20}{3} \wlin{15,40}{25,30}
\sof(325,10)
\flin{0,0}{15,10} \flin{15,10}{30,0}
\flin{30,50}{15,40} \flin{15,40}{0,50}
\flin{25,30}{40,40} \flin{25,20}{25,30} \flin{40,10}{25,20}
\zlin{15,10}{25,20} \wlin{15,40}{25,30}
\efig{Figure 6: background diagrams for $e^+e^-\to e^+e^-\nu_e\bar{\nu}_e$}
For the $\epl\epl\emn\emn$ final state, there are
32 bremsstrahlung diagrams, 4 conversion, 16 annihilation, and 32
multiperipheral ones. Of these, fig.7 depicts 21 only; the other ones
are obtained by permuting the outgoing positrons or electrons.

Since different final states select different sets of diagrams, an
inventory is made in tables \ref{tableone}-\ref{tablethree}.
All final states that have the same matrix element are grouped together.
Moreover, the number of Abelian diagrams ($N_a$) and of
non\--A\-be\-li\-an diagrams ($N_n$) is given.
We also indicate which of the signals
(1=(\ref{wpair}),\ldots,5=(\ref{singleznu}))
lead to the final state, and how many of the background diagrams of
figs.6 or 7 occur ($N_b$).
It is seen, that for the leptonic processes there are 15 different
matrix elements, for the semileptonic processes 10, and in the
purely hadronic case 7. Since differences in the latter two cases partly
arise from different quark coupling constants, the numbers of
{\em structurally\/} different matrix elements in these two cases
are 6 and 4, respectively.
\bfig(384,190)
\sof(0,140)
\flin{0,0}{20,10} \flin{20,10}{40,0} \glin{20,10}{20,20}{2}
\flin{40,25}{20,20} \flin{20,20}{10,25} \flin{10,25}{0,30}
\glin{10,25}{20,30}{2} \flin{35,35}{20,30} \flin{20,30}{25,45}
\sof(48,130)
\flin{0,0}{20,10} \flin{20,10}{40,0} \zlin{20,10}{20,20}
\flin{40,25}{20,20} \flin{20,20}{10,25} \flin{10,25}{0,30}
\glin{10,25}{20,30}{2} \flin{35,35}{20,30} \flin{20,30}{25,45}
\sof(96,140)
\flin{0,0}{15,10} \flin{15,10}{40,0} \glin{15,10}{15,30}{4}
\flin{40,40}{25,35} \flin{25,35}{15,30} \flin{15,30}{0,40}
\glin{25,35}{30,25}{2} \flin{40,10}{30,25} \flin{30,25}{45,20}
\sof(144,140)
\flin{0,0}{15,10} \flin{15,10}{40,0} \zlin{15,10}{15,30}
\flin{40,40}{25,35} \flin{25,35}{15,30} \flin{15,30}{0,40}
\glin{25,35}{30,25}{2} \flin{40,10}{30,25} \flin{30,25}{45,20}
\sof(192,140)
\flin{0,10}{10,15} \flin{10,15}{20,20} \flin{20,20}{40,15}
\glin{20,20}{20,30}{2} \flin{40,40}{20,30} \flin{20,30}{0,40}
\glin{10,15}{20,10}{2} \flin{30,-5}{20,10} \flin{20,10}{35,5}
\sof(240,140)
\flin{0,10}{10,15} \flin{10,15}{20,20} \flin{20,20}{40,15}
\zlin{20,20}{20,30} \flin{40,40}{20,30} \flin{20,30}{0,40}
\glin{10,15}{20,10}{2} \flin{30,-5}{20,10} \flin{20,10}{35,5}
\sof(288,140)
\flin{0,0}{15,10} \flin{15,10}{25,5} \flin{25,5}{40,0}
\glin{15,10}{15,30}{4} \flin{40,40}{15,30} \flin{15,30}{0,40}
\glin{25,5}{30,15}{2} \flin{45,20}{30,15} \flin{30,15}{40,30}
\sof(336,140)
\flin{0,0}{15,10} \flin{15,10}{25,5} \flin{25,5}{40,0}
\zlin{15,10}{15,30} \flin{40,40}{15,30} \flin{15,30}{0,40}
\glin{25,5}{30,15}{2} \flin{45,20}{30,15} \flin{30,15}{40,30}
\sof(30,80)
\flin{0,0}{10,15} \flin{10,15}{10,25} \flin{10,25}{0,40}
\glin{10,15}{20,10}{2} \flin{35,0}{20,10} \flin{20,10}{40,15}
\glin{10,25}{20,30}{2} \flin{40,25}{20,30} \flin{20,30}{35,40}
\sof(110,80)
\flin{0,0}{10,20} \flin{10,20}{0,40} \glin{10,20}{20,20}{2}
\flin{30,40}{25,30} \flin{25,30}{20,20} \flin{20,20}{30,0}
\glin{25,30}{30,20}{2} \flin{40,10}{30,20} \flin{30,20}{45,25}
\sof(170,80)
\flin{0,0}{10,20} \flin{10,20}{0,40} \zlin{10,20}{20,20}
\flin{30,40}{25,30} \flin{25,30}{20,20} \flin{20,20}{30,0}
\glin{25,30}{30,20}{2} \flin{40,10}{30,20} \flin{30,20}{45,25}
\sof(230,80)
\flin{0,0}{10,20} \flin{10,20}{0,40} \glin{10,20}{20,20}{2}
\flin{30,40}{20,20} \flin{20,20}{25,10} \flin{25,10}{30,0}
\glin{25,10}{30,20}{2} \flin{45,15}{30,20} \flin{30,20}{40,30}
\sof(290,80)
\flin{0,0}{10,20} \flin{10,20}{0,40} \zlin{10,20}{20,20}
\flin{30,40}{20,20} \flin{20,20}{25,10} \flin{25,10}{30,0}
\glin{25,10}{30,20}{2} \flin{45,15}{30,20} \flin{30,20}{40,30}
\sof(10,10)
\flin{0,0}{15,10} \flin{15,10}{30,0}
\flin{30,50}{15,40} \flin{15,40}{0,50}
\flin{40,40}{25,30} \flin{25,30}{25,20} \flin{25,20}{40,10}
\glin{15,10}{25,20}{2} \glin{15,40}{25,30}{2}
\sof(58,10)
\flin{0,0}{15,10} \flin{15,10}{30,0}
\flin{30,50}{15,40} \flin{15,40}{0,50}
\flin{40,40}{25,30} \flin{25,30}{25,20} \flin{25,20}{40,10}
\zlin{15,10}{25,20} \glin{15,40}{25,30}{2}
\sof(106,10)
\flin{0,0}{15,10} \flin{15,10}{30,0}
\flin{30,50}{15,40} \flin{15,40}{0,50}
\flin{40,40}{25,30} \flin{25,30}{25,20} \flin{25,20}{40,10}
\glin{15,10}{25,20}{2} \zlin{15,40}{25,30}
\sof(154,10)
\flin{0,0}{15,10} \flin{15,10}{30,0}
\flin{30,50}{15,40} \flin{15,40}{0,50}
\flin{40,40}{25,30} \flin{25,30}{25,20} \flin{25,20}{40,10}
\zlin{15,10}{25,20} \zlin{15,40}{25,30}
\sof(202,10)
\flin{0,0}{15,10} \flin{15,10}{30,0}
\flin{30,50}{15,40} \flin{15,40}{0,50}
\flin{25,30}{40,40} \flin{25,20}{25,30} \flin{40,10}{25,20}
\glin{15,10}{25,20}{2} \glin{15,40}{25,30}{2}
\sof(246,10)
\flin{0,0}{15,10} \flin{15,10}{30,0}
\flin{30,50}{15,40} \flin{15,40}{0,50}
\flin{25,30}{40,40} \flin{25,20}{25,30} \flin{40,10}{25,20}
\zlin{15,10}{25,20} \glin{15,40}{25,30}{2}
\sof(294,10)
\flin{0,0}{15,10} \flin{15,10}{30,0}
\flin{30,50}{15,40} \flin{15,40}{0,50}
\flin{25,30}{40,40} \flin{25,20}{25,30} \flin{40,10}{25,20}
\glin{15,10}{25,20}{2} \zlin{15,40}{25,30}
\sof(338,10)
\flin{0,0}{15,10} \flin{15,10}{30,0}
\flin{30,50}{15,40} \flin{15,40}{0,50}
\flin{25,30}{40,40} \flin{25,20}{25,30} \flin{40,10}{25,20}
\zlin{15,10}{25,20} \zlin{15,40}{25,30}
\efig{Figure 7: background diagrams for $e^+e^-\to e^+e^-e^+e^-$}

\section{The matrix elements}
The amplitudes receive contributions from Abelian and
non\--A\-be\-li\-an graphs, with distinct topological structure.
They are given in fig. 8.

In these so-called {\em generic\/} diagrams, all particles are assumed
to be outgoing: assigning two fermion legs to be the initial-state
fermions (by crossing), the actual Feynman diagrams are generated.
In the Abelian diagrams the charges of the fermions determine the
character of the two exchanged bosons, which may be $W^+$, $W^-$,
$Z$ or $\gamma$. In the non\--A\-be\-li\-an diagrams, two of the vector bosons
are fixed to be $W^+$ and $W^-$, and the third one can be $Z$ or
$\gamma$. In this way we avoid double-counting of diagrams.

The particles and antiparticles can each be assigned in six ways to
the external lines (in principle). This gives, for the Abelian graphs,
a maximum of 144 different diagrams, and at most 8 for the
non\--A\-be\-li\-an diagrams.

We evaluate the matrix element at the level of helicity amplitudes.
For a specific particle/antiparticle content of the final state,
the contributing diagrams are enumerated by permuting the assignments of
the external legs in the generic diagrams, and by relating the
charges of the fermions to those of the vector bosons $V_1$ and $V_2$.

\bfig(300,220)
\sof(20,20)
\flin{40,0}{0,20} \flin{0,20}{40,40}
\flin{40,60}{0,80} \flin{0,80}{0,120} \flin{0,120}{40,140}
\flin{40,160}{0,180} \flin{0,180}{40,200}
\wlin{0,20}{0,80} \wlin{0,120}{0,180}
\Text(45,0)[lc]{$p_6,\sigma$} \Text(45,40)[lc]{$p_5,\sigma$}
\Text(45,60)[lc]{$p_4,\rho$} \Text(45,140)[lc]{$p_3,\rho$}
\Text(45,160)[lc]{$p_2,\lambda$} \Text(45,200)[lc]{$p_1,\lambda$}
\Text(-5,50)[rc]{$V_1$} \Text(-5,150)[rc]{$V_2$}
\sof(200,40)
\flin{40,0}{0,20} \flin{0,20}{40,40}
\wlin{0,20}{0,80} \wlin{0,80}{40,80} \wlin{0,80}{0,140}
\flin{80,60}{40,80} \flin{40,80}{80,100}
\flin{40,120}{0,140} \flin{0,140}{40,160}
\Text(45,0)[lc]{$p_6,\sigma$}
\Text(45,40)[lc]{$p_5,\sigma$}
\Text(85,60)[lc]{$p_4,\rho$}
\Text(85,100)[lc]{$p_3,\rho$}
\Text(45,120)[lc]{$p_2,\lambda$}
\Text(45,160)[lc]{$p_1,\lambda$}
\Text(-5,50)[rc]{$W^-$}
\Text(-5,110)[rc]{$W^+$}
\Text(20,75)[tc]{$V$}
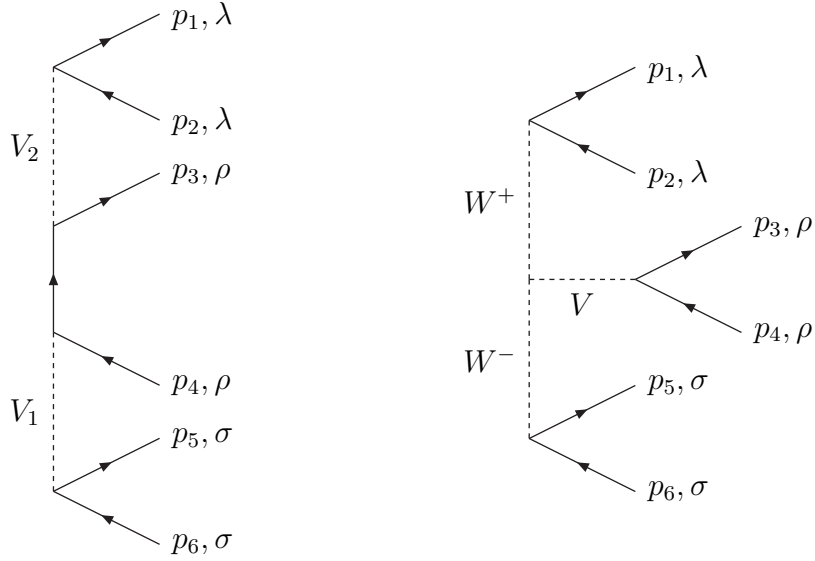
\efig{Figure 8: generic diagrams for four-fermion production.
The fermion momenta and helicities, and the bosons are indicated.
The bosons $V_{1,2}$ can be either $Z$, $W^{\pm}$, or $\gamma$;
$V$ can be either $Z$ or $\gamma$.}

Not all of these generated diagrams will contribute to each helicity
amplitude, as can be seen from the following expression for the
numerator of the abelian diagram of fig.8:
\bqa
\lefteqn{A(\lambda,\rho,\sigma;p_1,p_2,p_3,p_4,p_5,p_6) =}\nl
& = &\hphantom{\times\;\;}
 \ubar_{\lambda}(p_1)\gamma^{\mu}u_{\lambda}(p_2)\nl
& & \times\;\;
 \ubar_{\rho}(p_3)\gamma_{\mu}(\rlap/p_1+\rlap/p_2+\rlap/p_3)
\gamma_{\nu}u_{\rho}(p_4)\nl
& & \times\;\;
 \ubar_{\sigma}(p_5)\gamma^{\nu}u_{\sigma}(p_6)\;\;.
\eqa
Here we have disregarded the particle/antiparticle distinction
since it is already implied by the assignment of
the external momenta.
The helicity labels $\lambda,\rho,\sigma = \pm$ determine the helicity
of both external legs on a given fermion line. Using the
Weyl-van der Waerden formalism for helicity amplitudes \cite{spinors8}
(or, equivalently, the Dirac formalism of \cite{spinors7}),
the expression $A$ can easily be calculated. For instance, for
$\lambda=\rho=\sigma=1$ one finds
\bq
A(+,+,+;1,2,3,4,5,6) =
 -4\wws{3}{1}\ww{4}{6}\left[
 \wws{5}{1}\ww{2}{1}+\wws{5}{3}\ww{2}{3}\right]\;\;,
\label{afunc}
\eq
where the spinorial product is given, in terms of the momenta components,
by
\bq
\ww{k}{j} \equiv
 \left(p_j^1+ip_j^2\right)
 \left[{p_k^0-p_k^3 \over p_j^0-p_j^3}\right]^{1/2}
 \;\;-\;\;(k\leftrightarrow j)\;\;.
\eq
we denote the expression of \eqn{afunc} by $A_0(1,2,3,4,5,6)$.
All helicity combinations can be expressed in terms of $A_0$,
as follows:
\bqa
A(+++) = A_0(1,2,3,4,5,6) &\;\;\;& A(---) = A_0(1,2,3,4,5,6)^{\star}\;\;\;,\nl
A(-++) = A_0(2,1,3,4,5,6) &\;\;\;& A(+--) = A_0(2,1,3,4,5,6)^{\star}\;\;\;,\nl
A(++-) = A_0(1,2,3,4,6,5) &\;\;\;& A(--+) = A_0(1,2,3,4,6,5)^{\star}\;\;\;,\nl
A(-+-) = A_0(2,1,3,4,6,5) &\;\;\;& A(+-+) = A_0(2,1,3,4,6,5)^{\star}\;\;\;.
\eqa
The numerator in the non\--A\-be\-li\-an diagrams can also be written in
terms of the function $A$:
\bqa
\lefteqn{\ubar_{\lambda}(p_1)\gamma_{\alpha}u_{\lambda}(p_2)
       \;\ubar_{\rho}(p_3)\gamma_{\mu}u_{\rho}(p_4)
       \;\ubar_{\sigma}(p_5)\gamma_{\nu}u_{\sigma}(p_6)}\nl
\lefteqn{\times\;
 \left\{g^{\mu\alpha}(p_1+p_2)^{\nu}
       +g^{\alpha\nu}(p_5+p_6)^{\mu}
       +g^{\nu\mu}(p_3+p_4)^{\alpha}\right\}}\nl
& = & A(\lambda,\rho,\sigma;1,2,3,4,5,6) -
      A(\sigma,\rho,\lambda;5,6,3,4,1,2)\;\;.
\eqa
Thus, for massless fermions, every helicity amplitude consists of a sum of
very systematic, and relatively compact, expressions.
When the fermions acquire a non-zero mass, the spin states of the
external legs are not related so simply anymore. Additional helicity
amplitudes therefore occur, as well as extra terms in the
ones above, that are suppressed by one and two powers of the ratio
fermion mass/energy,
respectively. Also, new diagrams crop up, where massive
fermions are coupled to a Higgs boson. In these diagrams,
the unsuppressed helicity
amplitudes (that nonetheless contain the fermion mass in the
coupling constant to the Higgs) are precisely the ones that are
suppressed in the vector boson diagrams, and vice versa.
We may therefore conclude that the interference of the Higgs diagrams
with the vector boson diagrams will always be suppressed by a factor
proportional to the fermion mass squared.

Finally it should be noted that the vector boson propagators are implemented
in the form $q^2-M^2_V+{\em i}M_V\Gamma_V$, irrespective whether $q$ is
timelike or not. This only introduces a negligible error. A more refined
procedure with a $q^2$ dependent width is of course possible.

\section{The Monte Carlo}
In this section we describe the structure of {\tt EXALIBUR},
a Monte Carlo simulation program for the four-fermion production
processes discussed above. First, we shall review the basic ideas,
and then show their implementation in the actual simulation.\\

For the sake of simplicity let us concentrate on the
total cross section $\sigma$ for some process
\bqa
e^+(p_1)\;e^-(p_2)\;\;\rightarrow f_1(p_3)\;\bar{f}_2(p_4)
\;f_3(p_5)\;\bar{f}_4(p_6)\;\;.
\eqa
The main formula is
\bqa
\sigma= \int f(\vec\Phi)\;d\vec\Phi\;,
\eqa
where $f(\vec \Phi)$ denotes the matrix element squared
(any cut can be easily implemented by putting $f(\vec\Phi)=0$
in the unwanted region of the phase space) and
\bqa
d\vec \Phi= \prod_{i=3}^6 d^4p_i \;\delta(p^2_i)\;
\delta^4(p_1+p_2-p_3-p_4-p_5-p_6)
\eqa
is the 8-dimensional massless phase space integration element.

In order to reduce the variance of the integrand, and therefore the
Monte Carlo error, it is convenient to introduce an analytically
integrable function $g(\vec \Phi)$, called the {\em local density\/},
that exibits approximately the same peaking
behaviour of $f(\vec\Phi)$ and is {\em unitary\/}, that is,
a normalized probability density:
\bqa
\int g(\vec\Phi)\;d\vec\Phi= 1\;\;.
\eqa
By multiplying and dividing the integrand by $g(\vec \Phi)$, the cross
section can be rewritten as follows
\bqa
\sigma= \int w(\vec\Phi(\vec\rho))\;d\vec\rho
\eqa
where the new integrand
\bqa
w(\vec\Phi(\vec\rho)) &=&\frac{f(\vec\Phi)}{g(\vec\Phi)}
\eqa
is a smoother function of the new set of variables $\{\rho_i\}$ defined
by
\bqa
d\vec\rho = g(\vec\Phi)d\vec\Phi \nonumber \\
0<\rho_i<1\;\;
\eqa
so that the variance of $w(\vec\rho)$ is smaller than the variance of
$f(\vec\Phi)$.

When the peaking structure of the matrix element squared is very rich
(the worst case being $e^+ e^- \rightarrow e^+ e^- e^+ e^-$ with 144
different Feynman diagrams) one set of new integration variables
$\{ \rho_i\}$ can only describe well a limited number of peaks.
Therefore a multichannel approach is required in which
\bq
g(\vec \Phi) = \sum_{i= 1}^N \alpha_i\;g_i(\vec \Phi)\;\;,\;\;
\sum_{i= 1}^N \alpha_i = 1\;\;,\;\;
\int g_i(\vec \Phi) d \vec \Phi = 1\;\;,
\eq
and where every $g_i(\vec \Phi)$ describes a particular peaking structure
of $f(\vec\Phi)$. Note that the conditions on the $\alpha_i$ and
$g_i(\Phi)$ ensure unitarity of the algorithm, {\it i.e.\/}
probability is explicitly conserved at each step of the
algorithm, without additional normalization factors at any stage.

The numbers $\alpha_i$ are called {\em a-priori weights} and,
although their numerical values are in principle unimportant,
they can be used, in practice, to reduce the Monte Carlo error
\cite{multi10}.

In {\tt EXCALIBUR} we have dealt with the problem of the construction of
the $g_i(\vec\Phi)$ in a very modular and systematic way.
Firstly we have singled out all possible
kinematical diagrams \footnote{Here and in the following kinematical diagrams
are defined to be pictures that represent the various peaking structures.
Although they are inspired by the Feynman diagrams,
they should not interpreted further than that
they indicate which variables are most appropriate to a given
$g_i(\vec\Phi)$ \cite{multi10}.}
for every four-fermion final state. Secondly we have constructed all
building blocks (that is subroutines) necessary for the calculation.
Finally we have put them together to form the $g_i(\vec\Phi)$.\\

To illustrate the outlined procedure we shall treat in detail one particular
channel, that is the conversion channel with one massless and one
massive particle emitted.

Referring to the kinematical diagram in figure 9, a suitable choice for the
8 integration variables is
\begin{itemize}
\item the angle $\theta$ between $p_1$ and $p_3+p_4$ in the lab frame;
\item the azimuthal angle $\phi$ adjoint to $\theta$;
\item the decay angle $\theta_1$ of the particles 3 and 4 in the rest frame
of $(p_3+p_4)$, and its adjoint $\phi_1$;
\item the decay angle $\theta_2$ of the particles 5 and 6 in the rest frame
of $(p_5+p_6)$, and its adjoint $\phi_2$;
\item the squared invariant  masses $s_{34}=(p_3+p_4)^2$ and
$s_{56}=(p_5+p_6)^2$.
\end{itemize}
\bfig(160,160) \sof(10,20)
\flin{0,0}{40,20} \flin{40,20}{40,80} \flin{40,80}{0,100}
\glin{40,80}{100,90}{8} \wlin{40,20}{100,10}
\flin{140,70}{100,90} \flin{100,90}{140,110}
\flin{140,-10}{100,10} \flin{100,10}{140,30}
\CArc(40,80)(10,30,140) \Text(40,93)[b]{$\theta$}
\CArc(100,90)(10,-20,20) \Text(113,90)[l]{$\theta_1$}
\CArc(100,10)(10,-20,20) \Text(113,10)[l]{$\theta_2$}
\Text(90,7)[rt]{$M-i\Gamma$}
\Text(-5,0)[r]{$p_2$} \Text(-5,100)[r]{$p_1$}
\Text(145,110)[l]{$p_3$} \Text(145,70)[l]{$p_4$}
\Text(145,30)[l]{$p_5$}  \Text(145,-10)[l]{$p_6$}
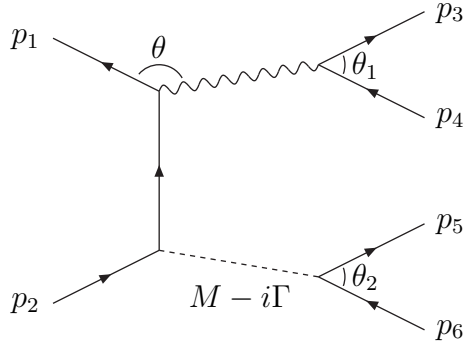
\efig{Figure 9: kinematical diagram representing a conversion with
emission of one massless and one massive particle.}

Furthermore we expect in the cross section $\cos \theta$ distributed like
\bq
\frac{1}{(a-\cos\theta)^\nu}\;\;\;,\;\;\;\nu \sim 1\;\;\;,\;\;\;
a= \frac{\sqrt s (E_3+E_4)-s_{34}}{\sqrt s |\vec p_3+\vec p_4|}\;\;,
\label{eq:tcha}
\eq
$s_{34}$ like
\bq
\frac{1}{s^{\nu_1}_{34}}\;\;\;,\;\;\;\nu_1 \sim 1\;\;,
\eq
and $s_{56}$ like
\bq
\frac{1}{{(s_{56}-M^2)}^2+M^2\Gamma^2}\;\;,
\eq
while all other distributions are expected to be more or less flat.
Therefore the phase space integration can be split into five
parts. Using
\bqa
1&=& \int ds_{34}\;d^4 P_{34}\;\delta^4(P_{34}-p_3-p_4)
\delta((p_3+p_4)^2-s_{34})\;\;,\nl
1&=& \int ds_{56}\;d^4 P_{56}\;\delta^4(P_{56}-p_5-p_6)\;
\delta((p_5+p_6)^2-s_{56})
\eqa
it follows
\bqa
\int d\vec \Phi &=& \underbrace{\int_{s_{min}}^s\!
\frac{ds_{34}}{s^{\nu_1}_{34}}
\cdot s^{\nu_1}_{34}}\nl
&&~~~~~~~~~\!{\em i}\nl
&&\underbrace{\int_0^{(\sqrt s- \sqrt s_{34})^2}
\!\!\frac{ds_{56}}{[(s_{56}-M^2)^2+M^2\Gamma^2]}
\cdot [(s_{56}-M^2)^2+M^2\Gamma^2]} \nl
&&~~~~~~~~~~~~~~~~~~~~~~~~~~~~~~~~~~~\,{\em ii}\nl
&\times& \underbrace{{\textstyle \frac{1}{8}}
\lambda^{\frac{1}{2}}\left(\frac{s_{34}}{s},
\frac{s_{56}}{s}\right) \int_{0}^{2 \pi} d\phi
\int_{-1}^{1}\frac{d \cos \theta}{(a-\cos\theta)^\nu}
\cdot (a-\cos\theta)^\nu}\nl
&&~~~~~~~~~~~~~~~~~~~~~~~~~~~~~~~\;{\em iii}~\nl
&\times&\underbrace{{\textstyle \frac{1}{8}}\int_0^{2\pi} d\phi_1 \int_{-1}^1
d \cos \theta_1}~~~\underbrace{{\textstyle \frac{1}{8}}
\int_0^{2\pi} d\phi_2 \int_{-1}^1 d \cos \theta_2}~~\\
&&~~~~~~~~~~~~~\!{\em iv}~~~~~~~~~~~~~~~~~~~~~~~~~~~~\,{\em v} \nonumber
\eqa
where $s_{min}$ is a minimum value for $s_{34}$ and
\bqa
\lambda(x,y)= 1+x^2+y^2-2x-2y-2xy\;\;.
\eqa
The first and the second contributions take care of the photon and
the massive boson propagators, the third one describes the {\em t-channel}
distribution of $\cos \theta$ while the last two integrals represent the
two body decays of the massless and massive boson respectively.
The new set of integration variables $\{\rho_i\}$ is
\bqa
s_{34} &=& [\rho_1 s^{(1-\nu_1)}+(1-\rho_1)s_{min}^{(1-\nu_1)}]^
  {\frac{1}{1-\nu_1}}\;\;, \nl
s_{56} &=& M^2+M\Gamma\;\tan\;[\rho_2(y^+-y^-)+y^-]\;\;,\nl
y^+ &=& \tan^{-1}
\left[\left(\sqrt s -\sqrt s_{34} )^2-M^2\right)/M\Gamma\right]\;\;,\nl
y^- &=& \tan^{-1}\left(-M/\Gamma\right) \;\;,\nl
\phi &=& 2 \pi\rho_3\;\;,\nl
\cos \theta &=& a-[\rho_4 (a-1)^{(1-\nu_1)}
+(1-\rho_4)(a+1)^{(1-\nu_1)}]^{\frac{1}{1-\nu_1}} \;\;,\nl
\phi_1&=& 2\pi\rho_5\;\;,\nl
\cos \theta_1 &=& 2\rho_6-1\;\;, \nl
\phi_2&=& 2\pi\rho_7\;\;\,\nl
\cos \theta_2 &=& 2\rho_8-1\;\;,
\label{eq:gen}
\eqa
so that it is possible to rewrite  
\bqa
\int d\vec \Phi &=& \int_0^1\!\!\prod_{i=1}^8\;d\rho_i
\underbrace{s_{34}^{\nu_1}}\;
\underbrace{[(s_{56}-M^2)^2+M^2\Gamma^2]} \nl
&& ~~~~~~~~~~~~~~g^{-1}_{\em i}~~~~~~~~~~~~~~g^{-1}_{\em ii}\nl
&\times& \underbrace{\textstyle{\frac{\pi}{4}}
\lambda^\frac{1}{2}\left(\frac{s_{34}}{s},
\frac{s_{56}}{s}\right) (a-\cos \theta)^\nu}\;\;\underbrace{\textstyle
{\frac{\pi}{2}}}\;\underbrace{\textstyle{\frac{\pi}{2}}} \nl
&& ~~~~~~~~~~~~~~g^{-1}_{\em iii}~~~~~~~~~~~~~~~~~
g^{-1}_{\em iv}~~~g^{-1}_{\em v}~~,
\eqa
Therefore the total local density can be written as a product of five
contributions
\bqa
g(\vec \Phi)= g_{\em i}\cdot g_{\em ii}\cdot g_{\em iii}\cdot
g_{\em iv}\cdot g_{\em v}\;\;.
\eqa
The advantage of this splitting is that every part of the algorithm can be
used again for other channels, both to generate distributions through
eqs.(\ref{eq:gen}) and to compute local densities. For example, using the
same ingredients, the channel in figure 10 can be easily built as follows
\bqa
\int d\vec \Phi &=& \underbrace{\int_0^s
\!\!\frac{ds_{56}}{[(s_{56}-M^2)^2+M^2\Gamma^2]}
\cdot [(s_{56}-M^2)^2+M^2\Gamma^2]}\nl
&&~~~~~~~~~~~~~~~~~~~~~~~~~~~~~~\!{\em ii}\nl
&\times&\underbrace{\int_{s_{56}}^s\!\frac{ds_{456}}{s^{\nu_1}_{456}}
\cdot s^{\nu_1}_{456}} \nl
&&~~~~~~~\,~{\em i}\nl
&\times& \underbrace{{\textstyle \frac{1}{8}}
  \lambda^{\frac{1}{2}}\left(\frac{s_3}{s},
\frac{s_{456}}{s}\right) \int_{0}^{2 \pi} d\phi
\int_{-1}^{1}\frac{d \cos \theta}{(1-\cos\theta)^\nu}
\cdot (1-\cos\theta)^\nu}\nl
&&~~~~~~~~~~~~~~~~~~~~~~~~~~~~~~~{\em iii}~\nl
&\times&\underbrace{{\textstyle \frac{1}{8}}\int_0^{2\pi} d\phi_1 \int_{-1}^1
d \cos \theta_1}~~~\underbrace{{\textstyle \frac{1}{8}}
\int_0^{2\pi} d\phi_2 \int_{-1}^1 d \cos \theta_2}~~\\
&&~~~~~~~~~~~~~{\em iv}~~~~~~~~~~~~~~~~~~~~~~~~~~~~\,{\em v} \nonumber
\eqa
where $s_3= p^2_3$, $s_{456}= (p_4+p_5+p_6)^2$ and ($\theta_1$,
$\phi_1$), ($\theta_2$, $\phi_2$) are the dacay angles of
$s_{456}$ and $s_{56}$.
\bfig(160,120) \sof(10,20)
\flin{0,0}{40,20} \flin{40,20}{70,5} \flin{70,5}{100,-10}
\glin{40,20}{40,60}{4}
\flin{100,90}{40,60} \flin{40,60}{0,80}
\wlin{70,5}{100,20}
\flin{140,0}{100,20} \flin{100,20}{140,40}
\Text(-5,80)[r]{$p_1$} \Text(-5,0)[r]{$p_2$}
\Text(105,90)[l]{$p_3$} \Text(105,-10)[l]{$p_4$}
\Text(145,40)[l]{$p_5$} \Text(145,0)[l]{$p_6$}
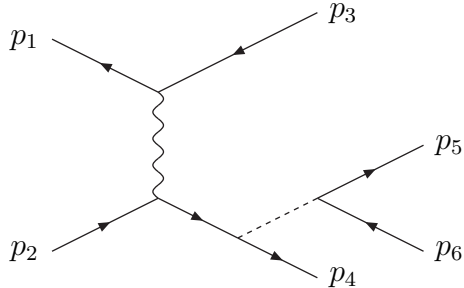
\efig{Figure 10: kinematical diagram representing a decaying massive boson
radiated off a final leg.}

This finishes our description of the event generation procedure. More details
can be found elsewhere \cite{bkp}. Finally, a brief remark is in order on
the collinear singularity in the multiperipheral and bremsstrahlung diagrams.
Consider an incoming electron with momentum $p^\mu$, which scatters over
a small angle into an outgoing momentum $q^\mu$, by emitting a
photon with virtuality t. We then have
\bqa
t= (p-q)^2= 2 m^2_e-2p^0q^0+2 |\vec p||\vec q| \cos \theta
\eqa
where $\theta$ is the scattering angle. For very small scattering angles,
the cross section will indeed be dominated by the Feymann diagrams with this
multiperipheral structure. If the electron is massive, there is no singularity,
in the sense that
\bqa
t &<& t_0\\
t_0&=& 2(m^2_e-p^0q^0+|\vec p||\vec q|) \sim -\frac{m^2_e(p_0-q_0)^2}{p_0q_0}
\eqa
where we have assumed $m_e<< q_0,p_0$. The total cross section is therefore
finite. For massless electrons this is not the case because
the collinear singularity can be reached. We may, however, mimic the dominant
effects of a nonzero mass by imposing, on the generated event, a cut
\bqa
\theta &>& \theta_0(p_0,q_0)\\
\theta_0(p_0,q_0)&=& \frac{m_e(p_0-q_0)}{p^0q^0}~.
\eqa
This cut leads, for massless electrons, to the same limit, $t_0$, on t that
would arise in the massive case. In our Monte Carlo approach, such a cut
can be easily applied to one or to both multiperipheral legs. In fact,
in the t-channel generation of $\theta$, the value of the exponent $\nu$ in eq.
\ref{eq:tcha} is arbitrary. Therefore, using $\nu<1$, $\cos \theta$ can be
generated in its full range [-1,1] even in the case of massless particles
and any final state dependent cut implemented in an event by event basis.
 Note, however, that this procedure only reproduces the leading
log result of the full, massive, cross section. For instance, subleading terms
due to spin-flip amplitudes are necessarily absent. On the other hand, the
angle $\theta_0$ is typically much smaller than any detector is likely to
cover, so the kinematics of the final state are, for all practical purposes,
those of the massless electron cross section.
\section{Results}
In this section we present some illustrative examples of the physics results
one can obtain with the event generator. The input parameters used are
the masses and widhts of the gauge bosons and the electroweak couplings
parametrized by $\alpha$ and $\sin^2\theta_W$. The masses and the widths are
taken to be independent parameters. The actual values in the program are
$\alpha= 1/129$, $\sin^2\theta_W= 0.23$, $M_W= 80.5$, $\Gamma_W= 2.3$,
$M_Z= 91.19$ and $\Gamma_Z= 2.5$ (all GeV). Note that, for simplicity, we
have obtained results for fixed, energy-independent widths. The change to
s-dependent widths is trivial by a simple modification of the boson
propagators computed for each Monte Carlo event.
There are three types of results.
The first type consists of checks against other calculations. The second type
is concerned with W-pair production at LEP 200 and somewhat higher energies.
The third type is related to high energy results for a future linear
$\epl \emn$ collider and illustrates signal/background issues for all five
signals.

As to the checks of the program, we first have verified that all toy model
matrix elements described in the appendix are, in fact, correctly computed
by the program. This serves to establish the consistency and reliability
of our approach. As to more physically meaningful comparisons,
results of the generator for the W-pair
signal have been obtained for a number of decreasing widths.
The extrapolated values compare well with the zero width analytic result
(table \ref{tablefour}).
In order to compare with a more complicated set of diagrams the semileptonic
channel $\mmn \numb u \dbar$ was chosen for which an analytic calculation
exists \cite{bardin}. Both the full calculation with all diagrams and the
signal calculation agree within the errors of the Monte Carlo calculation.

The LEP 200 results are divided up into leptonic, semileptonic and hadronic
final states.

As for the leptonic final states, results for two charged leptons and two
neutrinos are collected in table \ref{tablefive}. The first entry is the
W-pair signal, then the final states 1,6,16,20,18,2 and 8 of table
\ref{tableone} follow. They are calculated for all diagrams.
Reactions 16,18 and 20 are the non-interfering backgrounds.

Focussing on the results for a c.m.s. energy of 200 GeV, we see that taking
into account all diagrams for the final states $\epl \emn \nue \nueb$,
$\emn \nueb \num \mpl$, $\mpl \mmn \num \numb$, $\mmn \numb \nut \tpl$
 gives generally speaking
a larger cross section than the W-pair diagrams alone. One finds increases of
12, 3, 5, 0 \% respectively. The non-interfering backgrounds to the
$\epl \emn$ final state add up to 9.5 \% of the WW signal and for the
$\mpl \mmn$ or $\tpl \tmn$ final state to 9\%.

For completeness we note that there is also an invisible 4 neutrino cross
section. Without any cuts it has at 200 GeV the value $.8035~~10^{-1}$ pb, to
be compared with the total 2 neutrino cross section of $41.82$ pb.

The semileptonic states related to the W-pair signal are final states 1
and 5 of table \ref{tabletwo}. The results involving only the signal
diagrams and those containing all diagrams are listed in table \ref{tablesix}.
 Inclusions of all diagrams for the $\emn \nueb u \dbar$ final state gives
an increase of 4 \% above the WW signal at 200 GeV. When an invariant mass
cut is imposed on the quark pair, the effect does not change, while a
double invariant mass cut (provided the lepton-neutrino invariant mass can
be experimantally reconstructed) washes out any difference, as shown in
table \ref{tableseven}.
For the $\mmn \num u \dbar$ final state there is little difference between
the signal and the full calculation at LEP 200 energies.

The results for the 4 quark final states are given in table \ref{tableeight}.
The full calculation for the $u \ubar d \dbar$ final state gives an answer
3\% above the signal value, again at 200 GeV. For the $u \dbar s \cbar$
there is no difference between the signal and the full calculation.
When one adds all 4 jet events coming from the signal and does the same for
all non-interfering backgrounds, then the latter are 8\% of the real signal.
Note that those are non-interfering electroweak backgrounds and not
QCD backgrounds. These should be added as well, but this is beyond the
scope of the present paper.

The third type of results is for energies of 500, 1000 and 2000 GeV. It
should illustrate the possibilities of extracting the signal from a specific
final state, where the final state is produced by all diagrams. We take the
idealized case where one could measure any invariant mass combination
of the four fermions in the final state. When a charged or neutral fermion pair
 has an invariant mass $m\p{ij}$ in the interval $[M_V-2\Gamma, M_V+2\Gamma]$
for V= W or Z we identify the fermion pair as originating from the vector
boson V. In table \ref{tablenine} the $\emn \nueb \nue \epl$ final state
produced at 500 GeV is analyzed with respect to its vector boson content.
There are events which do not contain any mass combination giving rise
to a vector boson, but there are also some events containing apparently
three vector bosons.
It is now interesting to compare this analysis carried out on $\sit$
containing all diagrams with a similar analysis on $\sigma\p{i}$, the cross
section originating from the signal diagrams for process (i).
Tables \ref{tableten} - \ref{tablethirteen} do this for 4 different
final states. In table \ref{tableten} the $ \emn\nueb\nue\epl$ final
state is analyzed.
The cross section $\sit$ always is greater than the sum of all signals. This
demonstrates the relevance of the diagrams of fig. 6. Tables \ref{tableeleven}
- \ref{tablethirteen} consider $\emn\nue u \dbar$, $\epl \emn u \ubar$ and
$\nue \nueb u \ubar$ final states. Usually $\sit$ is greater than
$\sum_i\si\p{i}$, but not always. When one applies the idealized search
 algorithm to $\sit$ and to every $\sigma \p{i}$ both $\sit$
and $\sigma \p{i}$ are reduced, the former of course much more drastically.
The vector boson cross sections extracted from $\sit$ tend to be
greater than those originating from the signal cross section with a
possible exception at 2000 GeV.
The extracted WW cross sections are within the errors equal,
for $Z \epl \emn$ and
$Z \nu \nu$ this is also more or less the case at 1000 and 2000 GeV.

{}From these idealized examples it seems that search algorithms are possible
which extract the specific vector boson signal cross section. This causes a
 significant loss in statistics. Since most of the theoretical anomalous vector
boson coupling studies have been performed for non decaying vector boson final
 states, one has to take into account this reduction of statistics.
When one prefers to keep the higher statistics, then it seems recommendable
to study the effect of non-abelian couplings directly on the four-fermion
final states.

It should be noted that the energy dependences of the signals in tables
\ref{tableten} - \ref{tablethirteen} can differ from those in ref.
\cite{hagiwara}. The reason is that in the tables the cross sections always
have some cuts on the scattering angle, whereas in \cite{hagiwara} the full
phase space has been taken into account.
\section{Conclusions}
In this paper one strategy is chosen for the computation of all possible four
fermion final states. Whatever process is chosen the calculation proceeds
through the same series of steps. On one hand, this concerns the matrix
element which in principle contains all diagrams obtained by permutation
from two basic diagrams. On the other hand, it means that, for a specified
final state, the possible peaking structures are found and the phase space
is generated accordingly.
The algorithms for the generation of phase space are always made up of a
particular combination of elementary building blocks. The use of one procedure
for the matrix element and of a set of well tested (through model matrix
elements) algorithms for distributions of certain kinematical variables
makes the calculation transparent and less error prone. The neglect of
fermion masses renders the computation fast when one considers the often
large number of Feynman diagrams.

The four fermion final states offer a rich phenomenology: quite different
pure leptonic, semi leptonic and quark final states can be studied.
Moreover, the number of contributing diagrams to a specific
process varies a lot. In terms of produced vector bosons, which
subsequently decay, there are five signals which have been treated frequently
in the literature. For these the questions of signal and background
are discussed in this paper for a set of four fermion final states. It
is shown that sometimes the backgrounds are of the same order of magnitude
as radiative corrections and that certain experimental cuts can reduce the
backgrounds. Often both signal and background are then reduced, which causes
a loss in statistics.
Finally, some possible future extensions of our calculation or their
applications are listed
\begin{itemize}
\item[1)] certain studies, like effects of anomalous couplings have actually
 been performed for vector boson final states. In view of the above losses in
 statistics one may consider to build in directly anomalous couplings into
 the matrix element of section 3;
\item[2)] another possibility is to use the event generator for simulating
 non-interfering background events for Higgs searches;
\item[3)] when the QCD matrix elements are also taken into account, a full
analysis of four jet events in the region of vector boson production can be
performed;
\item[4)] radiative corrections to signals and backgrounds have not been
considered here. Some effects could be included in a further development of
the event generator. The most important one is the initial state radiation,
either in one loop or in a leading logarithmic approximation.
\end{itemize}
A last remark is in order. When one would like to study the precise
effects of untagged electrons or positrons one clearly needs massive fermions
in the matrix element and also an efficient algorithm for generating two
photon multiperipheral diagrams in the full phase space. This problem has been
dealt with in the literature in the past. It seems that one is unavoidably
led to time consuming event generators. Therefore an extension
of the present event generator to the exact massive fermion calculation
is bound to be quite involved and conducive to an appreciable loss in
program speed.
\section*{Acknowledgments}
Discussions with Dr. W. Beenakker are gratefully acknowledged.
We like to thank Dr. Bardin for providing us with numbers from his
analytical calculations.
We alkso thank the University of Wisconsin, in particular M. Livny et al.
from the Computer Science Department, for making available computing resources
at the Madison site through the Condor distributed batch system, and
Dr. R. van Dantzig for his help in using the Condor facilities.

\appendix
\section*{Appendix A: Model matrix elements and cross sections}
\setcounter{section}{1}
It is obvious that a Monte Carlo program such as {\tt EXCALIBUR} has
to be tested extensively before it can be considered reliable.
One of the most important kinds of tests is the Monte Carlo integration
of a matrix element that, at the one hand, displays some of the
salient features of the cross section, and,  on the other hand, has
an analytically known cross section. Note that this integral can be
computed for any allowed set of a-priori weights $\alpha_i$, that is,
either with the Monte Carlo channel $g_i(\vec{\Phi})$ that
gives the same peaking structure, or with another channel, or
with any combination of channels, and in each case the correct
cross section should be reproduced (with of course a larger error
if the appropriate channel is not included). Therefore, we give in this
appendix a list of such model matrix elements. These can be used either
for a user test of the Monte Carlo, and also (when
the relevant coupling constants are added) to get a rough idea
of the order of magnitude of the contribution of the
various kinematical regions to the total cross section.
In all cases, $p_{1,2}$ denote the incoming momenta, and
$p_{3,4,5,6}$ the produced fermion momenta. Each model matrix element
(squared and spin summed/averaged) is denoted by $T_i$, and the
corresponding total cross section is
\bq
\smod_i = {1\over(2s)(2\pi)^8}\int T_i\;d\vec{\Phi}\;\;.
\eq

\subsection{Conversion channels}
We distinguish the three cases of conversion of two photons, one photon
and one resonance and two resonances, respectively.
\begin{center}
\bmip{100}{100}\sof(20,15)
\lin{0,-10}{20,10} \lin{20,10}{20,40} \lin{20,40}{0,60}
\glin{20,10}{40,5}{4} \glin{20,40}{40,45}{4}
\lin{60,35}{40,45} \lin{40,45}{60,60}
\lin{60,-10}{40,5} \lin{40,5}{60,15}
\putk{-5,60}{1} \putk{-5,-10}{2}
\putp{65,60}{3} \putp{65,35}{4} \putp{65,15}{5} \putp{65,-10}{6}
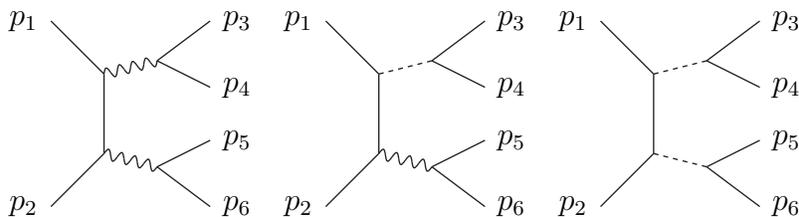
\emip{Figure 11: two-photon conversion}
\bmip{100}{100}\sof(20,15)
\lin{0,-10}{20,10} \lin{20,10}{20,40} \lin{20,40}{0,60}
\glin{20,10}{40,5}{4} \wlin{20,40}{40,45}
\lin{60,35}{40,45} \lin{40,45}{60,60}
\lin{60,-10}{40,5} \lin{40,5}{60,15}
\putk{-5,60}{1} \putk{-5,-10}{2}
\putp{65,60}{3} \putp{65,35}{4} \putp{65,15}{5} \putp{65,-10}{6}
\emip{Figure 12: photon-resonance conversion}
\bmip{100}{100}\sof(20,15)
\lin{0,-10}{20,10} \lin{20,10}{20,40} \lin{20,40}{0,60}
\wlin{20,10}{40,5} \wlin{20,40}{40,45}
\lin{60,35}{40,45} \lin{40,45}{60,60}
\lin{60,-10}{40,5} \lin{40,5}{60,15}
\putk{-5,60}{1} \putk{-5,-10}{2}
\putp{65,60}{3} \putp{65,35}{4} \putp{65,15}{5} \putp{65,-10}{6}
\emip{Figure 13: two-resonance conversion}
\end{center}

For these channels we define the function
\bqa
\tau(s_1,s_2)&=& s\left[\log\left(
{s-s_1-s_2+\lambda(s,s_1,s_2)\over
 s-s_1-s_2-\lambda(s,s_1,s_2)}\right)\right]^{-1} \;\;, \nl
\lambda(s,s_1,s_2)&=&\left((s-s_1-s_2)^2-4s_1s_2\right)^{1/2}\;\;.
\eqa
This function helps us to obtain analytical closed forms for the various
cross sections. Owing to the logarithm, it depends not too strongly
on $s_1$ and $s_2$.

\subsubsection{Two photons}
The kinematical situation is depicted in fig. 11. We have
$s_1= q^2_1= (p_3+p_4)^2$,  $s_2= q^2_2= (p_5+p_6)^2$ and
$Q^\mu= p_1^\mu-q_1^\mu$. In order to avoid singularities we impose
a lower bound on $s_1$ and $s_2$, and our model for $T_1$ becomes
\bq
T_1= {\tau(s_1,s_2) \over |Q^2|}~{\theta(s_{1,2}>s_0)\over s_1s_2}
\eq
and the cross section
\bq
\smod_1= F\left({\pi \over 2}\right)^3
\left[8\,\dilog{{w_0 \over w}}-2 {\pi^2 \over 3}
+4 \log^2\left({w_0 \over w}\right)\right]\;\;.
\eq
Here and in the following,
the step-function $\theta$ of an inequality is 1 if the inequality holds, and
0 otherwise. Furthermore, $w=\sqs$,
$w_i=\sqrt{s_i}$ (i= 0,1,2) and $F= {1\over (2s)(2\pi)^8}$.
\subsubsection{One photon and one resonance}
 One of  the photons now becomes a Breit-Wigner resonance with mass $m$ and
width $\Gamma$ (see fig. 12). We may now drop the lower limit on $s_1$:
\bq
T_2= {s\,\tau(s_1,s_2)\over|Q^2|}~{\theta(s_{2}>s_0)\over R(s_1)~s_2}\;\;,
\;\;R(s_1)= |s_1-m^2+im\Gamma|^2\;\;.
\eq
The total cross section is given by
\bqa
\smod_2 & = & F\left({\pi\over 2}\right)^3
{s\over 2im\Gamma}\left[A_2(z)-A_2(z\cjg)\right]\;\;,\;\;z=m^2+im\Gamma \nl
A_2(z) & = & 2\left[B_2(y)+B_2(-y)\right]~~~~~~~~~~~~~~~~~\,,\;\;y=\sqrt{z} \nl
B_2(y) & = & \lgn{{w-y\over w_0}}\lgn{{w-w_0-y\over-y}}\nl
& & -\dilog{{w-w_0-y\over w-y}}+\dilog{{-y\over w-y}}\;\;.
\eqa
Evaluation of this cross section relies on the evaluation of logarithms
and dilogarithms of complex arguments. Since their imaginary parts are
not infinitesimal, this poses no principal difficulty.
\subsubsection{Two resonances}
We use, for generality, two different masses $m_{1,2}$ and widths
$\Gamma_{1,2}$. Our model matrix element squared is now
\bq
T_3(s_1,s_2) =  {s^2\,\tau(s_1,s_2)\over R_1(s_1)R_2(s_2)}\;\;,\;\;
R_k(s_k)= |s_k-m^2_k+i\,m_k\Gamma_k|^2
\eq
and the total cross section $\smod_3$ is quite analogous to $\smod_2$;
\bqa
\smod_3 & = & F\left({\pi\over 2}\right)^3
{-s^2\over 4m_1m_2\Gamma_1\Gamma_2}\nl
&&\times\;\left[
A_3(z_1,z_2)-A_3(z_1,z_2\cjg)-A_3(z_1\cjg,z_2)+A_3(z_1\cjg,z_2\cjg)
\right]\;\;,\nl
z_k & = & m_k^2+im_k\Gamma_k\;\;,\nl
A_3(z_1,z_2)&=& B_3(y_1,y_2)+B_3(y_1,-y_2)+B_3(-y_1,y_2)
               +B_3(-y_1,-y_2)\;\;,\nl
y_k & = & \sqrt{z_k}\;\;, \nl
B_3(y_1,y_2) & = & \dilog{{y_2\over y_1+y_2-w}}-
\dilog{{y_2-w\over y_1+y_2-w}} +\lgn{1-{y_1+y_2\over w}}\nl
&\times& \left[ \lgn{1-{w\over y_1}}+
\lgn{{y_1\over y_1+y_2-w}}-\lgn{{y_1-w \over y_1+y_2-w}}\right] \nl
&-& \lgn{1- {w \over y_1}}\lgn{{y_2\over w}}-\lgn{1-{y_2\over w}}
\lgn{{y_1\over y_1+y_2-w}} \nl
&+&\lgn{-{y_2\over w}} \lgn{{y_1-w \over y_1+y_2-w}}\;\;.
\eqa
\subsection{Non-Abelian boson fusion channels}
Since the fusing bosons are in the t-channel, we neglect their $Q^2$
dependence if they are W or Z. This approximation is not so good if
$s>>m^2_{Z,W}$, but at LEP 200 energies it is quite accurate and, in any case,
we are allowed to propose any form for $T$, provided it can be checked
with our Monte Carlo.
\begin{center}
\bmip{100}{100}\sof(-5,10)
\lin{15,0}{40,20} \lin{40,20}{80,0} \lin{15,80}{40,60} \lin{40,60}{80,80}
\wlin{40,20}{45,40} \wlin{40,60}{45,40} \glin{45,40}{65,40}{4}
\lin{80,25}{65,40} \lin{65,40}{80,65}
\putk{10,80}{1} \putk{10,0}{2}
\putp{85,80}{5} \putp{85,65}{3} \putp{85,25}{4} \putp{85,0}{6}
\emip{Figure 14: two heavy bosons fusing into a photon}
\bmip{100}{100}\sof(-5,10)
\lin{15,0}{40,20} \lin{40,20}{80,0} \lin{15,80}{40,60} \lin{40,60}{80,80}
\wlin{40,20}{45,40} \wlin{40,60}{45,40} \wlin{45,40}{65,40}
\lin{80,25}{65,40} \lin{65,40}{80,65}
\putk{10,80}{1} \putk{10,0}{2}
\putp{85,80}{5} \putp{85,65}{3} \putp{85,25}{4} \putp{85,0}{6}
\emip{Figure 15: two heavy bosons fusing into a resonance}
\bmip{100}{100}\sof(-5,10)
\lin{15,0}{40,20} \lin{40,20}{80,0} \lin{15,80}{40,60} \lin{40,60}{80,80}
\wlin{40,20}{45,40} \glin{40,60}{45,40}{4} \wlin{45,40}{65,40}
\lin{80,25}{65,40} \lin{65,40}{80,65}
\putk{10,80}{1} \putk{10,0}{2}
\putp{85,80}{5} \putp{85,65}{3} \putp{85,25}{4} \putp{85,0}{6}
\Text(40,65)[b]{$\psi$}
\emip{Figure 16: boson and photon fusing into a resonance}
\end{center}
\subsubsection{Fusion of two heavy bosons into a photon}
The kinematics is given in fig. 14 and we write $s_1= (p_3+p_4)^2$.
Our model for this channel is
\bq
T_4= {\theta(s_1>s_0)\over s s_1}
\eq
and the cross section evaluates to
\bq
\smod_4=F\left({\pi\over 2}\right)^3{1\over 2}
\left[\left(1+{2 s_0\over s}\right) \log {s \over s_0}-{5 \over 2}
+{2s_0\over s}+{s^2_0\over 2 s^2}\right]\;\;.
\eq
\subsubsection{Two heavy bosons fusing into a resonance}
Replacing the photon of fig. 14 by a resonance, we get fig. 15 and
we have
\bq
T_5= {1 \over R(s_1)}\;\;,\;\;R(s_1)= |s_1-m^2+im\Gamma|^2
\eq
and
\bqa
\smod_5&=& F\left({\pi\over 2}\right)^3{s\over 4im\Gamma}
[A_5(z)-A_5(z\cjg)]\;\;,\;\;z=m^2+im\Gamma \nl
A_5(z)&=& \left(1-{z^2\over s^2}\right)\lgn{1-{s\over z}}
-{5\over 2}-{z\over s}
+{2z\over s} \dilog{{s\over z}}\;\;.
\eqa
\subsubsection{One photon and one heavy boson fusing into a resonance}
To avoid collinear singularities, we now have to impose an angular cut.
In fig. 16, let $\psi$ denote the angle between $\vec p_1$ and $\vec p_5$
in the lab frame. We impose a symmetrical cut $\psi_0<\psi<\pi- \psi_0$.
Modelling the (complicated) photon propagator by a simple angular dependence,
we may put
\bq
T_6= {\theta(\psi_0<\psi<\pi-\psi_0) \over 1-\cos \psi}~T_5\;\;.
\eq
So the cross section is obtained immediately:
\bq
\smod_6= {1 \over 2} \log \left({1+\cos\psi_0 \over 1-\cos\psi_0}\right)
\smod_5\;\;.
\eq
\subsection{Annihilation channels}
Since the total energy is fixed, it is immaterial whether the boson in
which the $e^+ e^-$ annihilate is a photon or a resonance (of course,
for the above mentioned order of magnitude estimates one may want to put in the
correct energy dependence).
\begin{center}
\bmip{110}{100}\sof(15,10)
\lin{0,0}{20,40} \lin{20,40}{0,80}
\wlin{20,40}{40,40} \wlin{40,40}{55,55} \wlin{40,40}{55,25}
\lin{55,55}{70,80} \lin{55,55}{75,50} \lin{55,25}{75,30} \lin{55,25}{70,0}
\putk{-5,0}{2} \putk{-5,80}{1}
\putp{75,80}{3} \putp{80,50}{4} \putp{80,30}{5} \putp{75,0}{6}
\emip{Figure 17: annihilation into two resonances}
\bmip{110}{100}\sof(15,10)
\lin{0,0}{20,40} \lin{20,40}{0,80}
\wlin{20,40}{40,40} \glin{40,40}{55,55}{4} \wlin{40,40}{55,25}
\lin{55,55}{70,80} \lin{55,55}{75,50} \lin{55,25}{75,30} \lin{55,25}{70,0}
\putk{-5,0}{2} \putk{-5,80}{1}
\putp{75,80}{3} \putp{80,50}{4} \putp{80,30}{5} \putp{75,0}{6}
\emip{Figure 18: annihilation into a photon and a resonance}
\end{center}
\begin{center}
\bmip{120}{100} \sof(20,10)
\lin{0,0}{20,40} \lin{20,40}{0,80} \wlin{20,40}{40,40}
\lin{40,40}{80,80} \lin{40,40}{80,0}
\glin{50,50}{70,55}{4} \lin{70,55}{80,65} \lin{70,55}{80,35}
\putk{-5,80}{1} \putk{-5,0}{2}
\putp{85,80}{5} \putp{85,65}{3} \putp{85,35}{4} \putp{85,0}{6}
\emip{Figure 19: annihilation into a photon and two fermions}
\bmip{120}{100} \sof(20,10)
\lin{0,0}{20,40} \lin{20,40}{0,80} \wlin{20,40}{40,40}
\lin{40,40}{80,80} \lin{40,40}{80,0}
\wlin{50,50}{70,55} \lin{70,55}{80,65} \lin{70,55}{80,35}
\putk{-5,80}{1} \putk{-5,0}{2}
\putp{85,80}{5} \putp{85,65}{3} \putp{85,35}{4} \putp{85,0}{6}
\emip{Figure 20: annihilation into a resonance and two fermions}
\end{center}

\subsubsection{Annihilation into two resonances}
The kinematics is given in fig. 17 and, as before, $s_i= q^2_i$.
We now take
\bq
T_7= \chi(s_1,s_2)~{s^2 \over R_1(s_1)R_2(s_2)}\;\;,
\eq
where we have introduced a factor analogous to $\tau(s_1,s_2)$:
\bq
\chi(s_1,s_2)= {s\over \lambda(s,s_1,s_2)}~\theta(w_1+w_2<\sqrt{s_m})\;\;,
\;\;s_m<s\;\;.
\eq
The cutoff is intended to avoid the singularity in $\lambda$ at $w_1+w_2= w$.
The integration is very similar to that of $T_3$:
\bqa
\smod_7&=& F\left({\pi\over 2}\right)^3{-s^2\over 4m_1m_2\Gamma_1\Gamma_2} \nl
&& \times \left[A_7(z_1,z_2)-A_7(z_1,z_2\cjg)-A_7(z_1\cjg,z_2)+
A_7(z_1\cjg,z_2\cjg)\right]\;\;,\nl
A_7(z_1,z_2) &=& \left[A_3(z_1,z_2)\right]_{s \to s_m}\;\;.
\eqa
\subsubsection{Annihilation into a photon and a resonance}
This channel can, strictly speaking, not occur in $e^+ e^-$ collisions, but
we include it for completeness, with an eye to, for instance, electron-quark
scattering. Fig. 18 gives the kinematics.

We put
\bq
T_8= \chi(s_1,s_2)~{s~\theta(s_1>0) \over s_1 R(s_2)}
\eq
so that we have
\bqa
\smod_8&=& F\left({\pi\over 2}\right)^3{s\over 2im\Gamma}
\left[A_8(z)-A_8(z\cjg)\right]\;\;, \nl
A_8(z)&=&\left[A_2(z)\right]_{s\to s_m}\;\;.
\eqa
\subsubsection{Annihilation into two fermions and a photon}
This situation is given in fig. 19. A cut on $s_1=(p_3+p_4)^2=q^2_1$ now
automatically imposes a cut on $s_2=(p_5+q_1)^2$, and we may use
\bq
T_9= {1 \over s_1s_2}~\theta(s_1>s_0)
\eq
leading to
\bq
\smod_9= F\left({\pi\over 2}\right)^3
\left[ {1 \over 2}\left(\log {s \over s_0}\right)^2-\left(2+{s_0 \over s}
\right) \log {s \over s_0}+3\left(1-{s_0 \over s} \right) \right]\;\;.
\eq
\subsubsection{Annihilation into two fermions and a resonance}
Replacing the photon of fig. 19 by a resonance, we obtain fig. 20.
We have the following model:
\bq
T_{10}= {s \over s_2R(s_1)}
\eq
and
\bqa
\smod_{10}&=& F\left({\pi\over 2}\right)^3{s\over 2im\Gamma}
\left[A_{10}(z)-A_{10}(z\cjg)\right]\;\; \nl
A_{10}(z) &=& 3-2\left(1-{z \over s}\right) \lgn{1-{s \over z}}
-\left(1+{z \over s} \right) \dilog{{s\over z}}\;\;.
\eqa
\subsection{Final state bremsstrahlung channels}
In all these cases, a boson is exchanged in the t channel and
a fermion pair radiated off an outgoing particle.
\begin{center}
\bmip{120}{100} \sof(20,10)
\lin{0 ,0 }{20,20} \lin{20,20}{80,0 } \lin{0 ,80}{20,60} \lin{20,60}{80,80}
\wlin{20,20}{20,60} \glin{35,15}{60,40}{6} \lin{60,40}{80,55}\lin{60,40}{80,25}
\putp{85,80}{5} \putp{85,55}{3} \putp{85,25}{4} \putp{85,0 }{6}
\putk{-5,80}{1} \putk{-5,0 }{2}
\emip{Figure 21: photon bremsstrahlung in heavy boson exchange}
\bmip{120}{100} \sof(20,10)
\lin{0 ,0 }{20,20} \lin{20,20}{80,0 } \lin{0 ,80}{20,60} \lin{20,60}{80,80}
\wlin{20,20}{20,60} \wlin{35,15}{60,40} \lin{60,40}{80,55} \lin{60,40}{80,25}
\putp{85,80}{5} \putp{85,55}{3} \putp{85,25}{4} \putp{85,0 }{6}
\putk{-5,80}{1} \putk{-5,0 }{2}
\emip{Figure 22: resonance bremsstrahlung in heavy boson exchange}
\end{center}
\begin{center}
\bmip{120}{100} \sof(20,10)
\lin{0 ,0 }{20,20} \lin{20,20}{80,0 } \lin{0 ,80}{20,60} \lin{20,60}{80,80}
\glin{20,20}{20,60}{6}\glin{35,15}{60,40}{6}\lin{60,40}{80,55}\lin{60,40}{80,25}
\putp{85,80}{5} \putp{85,55}{3} \putp{85,25}{4} \putp{85,0 }{6}
\putk{-5,80}{1} \putk{-5,0 }{2}
\emip{Figure 23: photon bremsstrahlung in photon exchange}
\bmip{120}{100} \sof(20,10)
\lin{0 ,0 }{20,20} \lin{20,20}{80,0 } \lin{0 ,80}{20,60} \lin{20,60}{80,80}
\glin{20,20}{20,60}{6}\wlin{35,15}{60,40} \lin{60,40}{80,55} \lin{60,40}{80,25}
\putp{85,80}{5} \putp{85,55}{3} \putp{85,25}{4} \putp{85,0 }{6}
\putk{-5,80}{1} \putk{-5,0 }{2}
\emip{Figure 24: resonance bremsstrahlung in photon exchange}
\end{center}
\subsubsection{Heavy boson exchange}
If the radiated fermion pair comes from a photon, the situation is given in
fig. 21. Since there is (by the definition of our model) no angular
dependence arising from the boson exchange, we simply have
\bq
T_{11}= T_9\;\;,\;\;\smod_{11}= \smod_9\;\;.
\eq
On the other hand, for resonance bremsstrahlung, the appropriate model for
fig. 22 is again one that we have already mentioned:
\bq
T_{12}= T_{10}\;\;,\;\;\smod_{12}= \smod_{10}\;\;.
\eq
\subsubsection{Photon exchange}
Now, we have a nontrivial angular dependence, which we again simply model by
the scattering angle $\psi$ between $\vec p_1$ and $\vec p_5$, in the lab
frame. For photon bremsstrahlung (fig. 23) we have a simple modification
of $T_{11}$;
\bqa
T_{13}&= &T_{11}~{ \theta(\psi_0<\psi<\pi-\psi_0) \over 1-\cos\psi} \nl
\smod_{13} &=& {1 \over 2} \log \left( {1+\cos \psi_0 \over
1-\cos \psi_0} \right) \smod_{11}\;\;.
\eqa
Similarly, for the resonance bremsstrahlung of fig. 24:
\bqa
T_{14}&= &T_{12}~{\theta(\psi_0<\psi<\pi-\psi_0) \over 1-\cos\psi} \nl
\smod_{14} &=& {1 \over 2} \log \left( {1+\cos \psi_0 \over
1-\cos \psi_0} \right) \smod_{12}\;\;.
\eqa
\subsection{Initial state bremsstrahlung}
The fermion pair is radiated off an initial leg. This complicates
matters somewhat.
\begin{center}
\bmip{120}{100} \sof(20,10)
\lin{0,0}{40,20} \lin{40,20}{80,0} \lin{0,60}{40,40} \lin{40,40}{80,55}
\wlin{40,20}{40,40} \glin{20,50}{40,60}{5}\lin{40,60}{60,80}\lin{40,60}{65,65}
\putk{-5,60}{1} \putk{-5,0}{2}
\putp{85,55}{5} \putp{85,0}{6} \putp{65,80}{3} \putp{70,65}{4}
\emip{Figure 25: photon bremsstrahlung in heavy boson exchange}
\bmip{120}{100} \sof(20,10)
\lin{0,0}{40,20} \lin{40,20}{80,0} \lin{0,60}{40,40} \lin{40,40}{80,55}
\wlin{40,20}{40,40} \wlin{20,50}{40,60} \lin{40,60}{60,80}\lin{40,60}{65,65}
\putk{-5,60}{1} \putk{-5,0}{2}
\putp{85,55}{5} \putp{85,0}{6} \putp{65,80}{3} \putp{70,65}{4}
\emip{Figure 26: resonance bremsstrahlung in heavy boson exchange}
\end{center}

\begin{center}
\bmip{120}{100} \sof(20,10)
\lin{0,0}{40,20} \lin{40,20}{80,0} \lin{0,60}{40,40} \lin{40,40}{80,55}
\glin{40,20}{40,40}{4}\glin{20,50}{40,60}{5}\lin{40,60}{60,80}\lin{40,60}{65,65}
\putk{-5,60}{1} \putk{-5,0}{2}
\putp{85,55}{5} \putp{85,0}{6} \putp{65,80}{3} \putp{70,65}{4}
\emip{Figure 27: photon bremsstrahlung in photon exchange}
\bmip{120}{100} \sof(20,10)
\lin{0,0}{40,20} \lin{40,20}{80,0} \lin{0,60}{40,40} \lin{40,40}{80,55}
\glin{40,20}{40,40}{4} \wlin{20,50}{40,60} \lin{40,60}{60,80}\lin{40,60}{65,65}
\putk{-5,60}{1} \putk{-5,0}{2}
\putp{85,55}{5} \putp{85,0}{6} \putp{65,80}{3} \putp{70,65}{4}
\emip{Figure 28: resonance bremsstrahlung in photon exchange}
\end{center}
\subsubsection{Heavy boson exchange}
Since the heavy boson propagator is again assumed to give a constant angular
distribution, these channels look somewhat like the conversion channels.
We define, for fig. 25 ($s_1=q^2_1=(p_3+p_4)^2,~s_2= (p_5+p_6)^2,~
Q^\mu=p^{\mu}_1-q^{\mu}_1$)
\bq
T_{15}= {\tau(s_1,s_2)~\theta(s_1>s_0) \over s s_1 |Q^2|}
\eq
which leads to
\bq
\smod_{15}= F\left({\pi\over 2}\right)^3 \left[\log{s\over s_0}
+1-{s_0 \over s}-4\left(1-{w_0\over w}\right)\right]\;\;.
\eq
For fig. 26, we have, in analogy:
\bq
T_{16}= {\tau(s_1,s_2) \over |Q^2| R(s_1)}
\eq
and
\bqa
\smod_{16}&=& F\left({\pi\over 2}\right)^3{1\over 2im\Gamma}
\left[A_{16}(z)-A_{16}(z\cjg)\right] \nl
A_{16}(z)&=& B_{16}(y)+B_{16}(-y) \nl
B_{16}(y)&=&(w-y)^2 \log \left(1-{w\over y}\right)-{3 w^2\over y}+wy\;\;.
\eqa
\subsubsection{Photon exchange}
For these channels, the situation is more complicated.
It is straightforward to define the scattering angle $\psi$ between
$\vec p_2$ and $\vec p_6$ not in the lab frame, but in the rest frame of
$\vec p_5+\vec p_6$, but a simple cut on this angle translates to a very
complicated one in the lab frame. We therefore again rely on our freedom to
postulate any reasonable behaviour, and simply model the photon propagator
such that it is simple in terms of this new $\psi$, and non singular upon
integration. At small scattering angles, this is actually not a bad
approximation to the real situation.
For photon bremsstrahlung (fig. 27) we therefore propose
\bq
T_{17}= T_{15}~{1 \over (1-\cos \psi)^\alpha}\;\;,\;\;\alpha<1\;\;.
\eq
So that
\bq
\smod_{17}= \left({2^{-\alpha}\over 1-\alpha}\right) \smod_{15}\;\;.
\eq
Similarly, for the resonance bremsstrahlung of fig. 28, we use
\bqa
T_{18}&=&T_{16}~{1 \over (1-\cos\psi)^\alpha}\;\;, \nl
\smod_{18}&=&\left({2^{-\alpha}\over 1-\alpha}\right) \smod_{16}\;\;.
\eqa
Note that the algorithms used in the Monte Carlo allow us to actually
generarate such non-singular angular distributions without restriction on the
scattering angles.
\subsection{Multiperipheral scattering}
We can distinguish according to the type of exchanged bosons
\begin{center}
\bmip{110}{100} \sof(15,10)
\lin{0,5}{20,20} \lin{20,20}{60,0} \lin{0,85}{20,70} \lin{20,70}{60,90}
\wlin{20,20}{30,35} \wlin{30,55}{20,70}
\lin{70,20}{30,35} \lin{30,35}{30,55} \lin{30,55}{70,70}
\putk{-5,85}{1} \putk{-5,5}{2}
\putp{65,90}{3} \putp{65,0}{6} \putp{75,70}{4} \putp{75,20}{5}
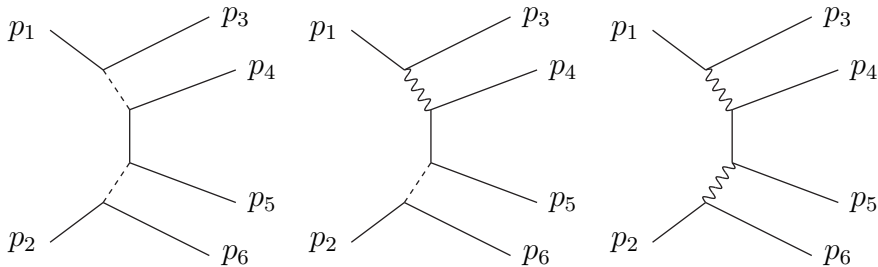
\emip{Figure 29: Two heavy bosons fusing into fermions}
\bmip{110}{100} \sof(15,10)
\lin{0,5}{20,20} \lin{20,20}{60,0} \lin{0,85}{20,70} \lin{20,70}{60,90}
\wlin{20,20}{30,35} \glin{30,55}{20,70}{4}
\lin{70,20}{30,35} \lin{30,35}{30,55} \lin{30,55}{70,70}
\putk{-5,85}{1} \putk{-5,5}{2}
\putp{65,90}{3} \putp{65,0}{6} \putp{75,70}{4} \putp{75,20}{5}
\emip{Figure 30: photon and heavy boson fusing into fermions}
\bmip{110}{100} \sof(15,10)
\lin{0,5}{20,20} \lin{20,20}{60,0} \lin{0,85}{20,70} \lin{20,70}{60,90}
\glin{20,20}{30,35}{4} \glin{30,55}{20,70}{4}
\lin{70,20}{30,35} \lin{30,35}{30,55} \lin{30,55}{70,70}
\putk{-5,85}{1} \putk{-5,5}{2}
\putp{65,90}{3} \putp{65,0}{6} \putp{75,70}{4} \putp{75,20}{5}
\emip{Figure 31: Two photons fusing into fermions}
\end{center}
We shall neglect the boson propagators for simplicity. Again we are
free to do this. We have $s_1=q^2_1=(p_3+p_4)^2,~s_2=
(p_5+p_6)^2,~Q^\mu=p^{\mu}_1-q^{\mu}_1$.
For the exchange of two heavy bosons we have simply the model
\bq
T_{19}= {\tau(s_1,s_2) \over s^2|Q|^2}\;\;,\;\;\smod_{19}=
F\left({\pi\over 2}\right)^3{1\over 6}\;\;.
\eq
Now for each photon, we apply the trick mentioned above.
Let $\psi_1$ be the angle $(\vec p_1, \vec p_3)$ in the $(\vec p_3+
\vec p_4)$ rest frame and $\psi_2$ that of $(\vec p_2, \vec p_6)$ in the
$(\vec p_5+\vec p_6)$ rest frame. Then, for fig. 30 and 31 respectively,
 we write
\bqa
T_{20}&=&T_{19}~{1 \over (1-\cos\psi_1)^\alpha}\;\;,\;\;
\smod_{20}=\left({2^{-\alpha}\over 1-\alpha}\right) \smod_{19} \nl
T_{21}&=&T_{20}~{1 \over (1-\cos\psi_2)^\alpha}\;\;,\;\;
\smod_{21}=\left({2^{-\alpha}\over 1-\alpha}\right) \smod_{20} \;\;.
\eqa
This finishes our list of matrix element models.

\begin{table}[h]
\begin{center}
\begin{tabular}{|l|l|c|c|c|c|l|} \hline\hline
label & final state & $N_a$ & $N_n$ & total & $N_b$ & signals \\
\hline
1 & $\epl\emn\nue\nueb$ & 48 &  8 &  56 & 12 & 1 2 3 4 5\\
\hline
2 & $\emn\nueb\num\mpl$ & 14 &  4 &  18 &  3 & 1 3 \\
3 & $\emn\nueb\nut\tpl$ & & & & & \\
4 & $\nue\epl\mmn\numb$ & & & & & \\
5 & $\nue\epl\tmn\nutb$ & & & & & \\
\hline
6 & $\mpl\mmn\num\numb$ & 17 & 2 & 19 & 0 & 1 2 \\
7 & $\tpl\tmn\nut\nutb$ & & & & & \\
\hline
8 & $\mmn\numb\nut\tpl$ & 7 & 2 & 9 & 0 & 1 \\
9 & $\tmn\nutb\num\mpl$ & & & & & \\
\hline
10 & $\epl\emn\epl\emn$ & 144 & 0 & 144 & 84 & 2 4 \\
\hline
11 & $\epl\emn\mpl\mmn$ & 48 & 0 & 48 & 26 & 2 4 \\
12 & $\epl\emn\tpl\tmn$ & & & & & \\
\hline
13 & $\mpl\mmn\mpl\mmn$ & 48 & 0 & 48 & 20 & 2 \\
14 & $\tpl\tmn\tpl\tmn$ & & & & & \\
\hline
15 & $\mpl\mmn\tpl\tmn$ & 24 & 0 & 24 & 10 & 2 \\
\hline
16 & $\epl\emn\num\numb$ & 20 & 0 & 20 & 2 & 2 4 \\
17 & $\epl\emn\nut\nutb$ & & & & & \\
\hline
18 & $\nue\nueb\mpl\mmn$ & 17 & 2 & 19 & 4 & 2 5 \\
19 & $\nue\nueb\tpl\tmn$ & & & & & \\
\hline
20 & $\nut\nutb\mpl\mmn$ & 10 & 0 & 10 & 0 & 2 \\
21 & $\num\numb\tpl\tmn$ & & & & & \\
\hline
22 & $\nue\nueb\nue\nueb$ & 32 & 4 & 36 & 4 & 2 5 \\
\hline
23 & $\nue\nueb\num\numb$ & 11 & 1 & 12 & 1 & 2 \\
24 & $\nue\nueb\nut\nutb$ & & & & & \\
\hline
25 & $\num\numb\num\numb$ & 12 & 0 & 12 & 0 & 2 5 \\
26 & $\nut\nutb\nut\nutb$ & & & & & \\
\hline
27 & $\num\numb\nut\nutb$ & 6 & 0 & 6 & 0 & 2 \\
\hline \hline
\end{tabular}
\end{center}
\caption[.]{leptonic four-fermion final states in $\epl\emn$ collisions.}
\label{tableone}
\end{table}

\newpage
\begin{table}[h]
\begin{center}
\begin{tabular}{|l|l|c|c|c|c|l|} \hline\hline
label & final state & $N_a$ & $N_n$ & total & $N_b$ & signals \\
\hline
1 & \ru{$\emn\nueb u\dbar$} & 16 & 4 & 20 & 4 & 1 3 \\
2 & $\emn\nueb c\sbar$ & & & & &  \\
3 & $\nue\epl d\ubar$  & & & & &  \\
4 & $\nue\epl s\cbar$ & & & & &  \\
\hline
5 & \ru{$\mmn\numb u\dbar$} & 8 & 2 & 10 & 0 & 1 \\
6 & $\mmn\numb c\sbar$ & & & & & \\
7 & $\mpl\num d\ubar$ & & & & & \\
8 & $\mpl\num s\cbar$ & & & & & \\
9 & $\tmn\nutb u\dbar$ & & & & & \\
10 & $\tmn\nutb c\sbar$ & & & & & \\
11 & $\tpl\nut d\ubar$ & & & & & \\
12 & $\tpl\nut s\cbar$ & & & & & \\
\hline
13 & $\epl\emn u\ubar$ & 48 & 0 & 48 & 26 & 2 4 \\
14 & $\epl\emn c\cbar$ & & & & & \\
\hline
15 &\ru{$\epl\emn d\dbar$} & 48 & 0 & 48 & 26 & 2 4 \\
16 & $\epl\emn s\sbar$ & & & & & \\
17 & $\epl\emn b\bbar$ & & & & & \\
\hline
18 & $\mpl\mmn u\ubar$ & 24 & 0 & 24 & 10 & 2   \\
19 & $\mpl\mmn c\cbar$ & & & & & \\
20 & $\tpl\tmn u\ubar$ & & & & & \\
21 & $\tpl\tmn c\cbar$ & & & & & \\
\hline \hline
\end{tabular}
\end{center}
\caption[.]{semileptonic four-fermion final states in $\epl\emn$ collisions
 {\it (continued on next page)}.}
\label{tabletwo}
\end{table}

\newpage
\setcounter{table}{1}
\begin{table}[h]
\begin{center}
\begin{tabular}{|l|l|c|c|c|c|l|} \hline\hline
label & final state & $N_a$ & $N_n$ & total & $N_b$ & signals \\
\hline
22 &\ru{$\mpl\mmn d\dbar$} & 24 & 0 & 24 & 10 & 2   \\
23 & $\mpl\mmn s\sbar$  & & & & & \\
24 & $\mpl\mmn b\bbar$  & & & & & \\
25 & $\tpl\tmn d\dbar$  & & & & & \\
26 & $\tpl\tmn s\sbar$  & & & & & \\
27 & $\tpl\tmn b\bbar$  & & & & & \\
\hline
28 & $\nue\nueb u\ubar$ & 17 & 2 & 19 & 4 & 2 5 \\
29 & $\nue\nueb c\cbar$ & & & & & \\
\hline
30 & \ru{$\nue\nueb d\dbar$} & 17 & 2 & 19 & 4 & 2 5 \\
31 & $\nue\nueb s\sbar$ & & & & & \\
32 & $\nue\nueb b\bbar$ & & & & & \\
\hline
33 & $\num\numb u\ubar$ & 10 & 0 & 10 & 0 & 2 \\
34 & $\num\numb c\cbar$ & & & & & \\
35 & $\nut\nutb u\ubar$ & & & & & \\
36 & $\nut\nutb c\cbar$ & & & & & \\
\hline
37 & \ru{$\num\numb d\dbar$} & 10 & 0 & 10 & 0 & 2 \\
38 & $\num\numb s\sbar$ & & & & & \\
39 & $\num\numb b\bbar$ & & & & & \\
40 & $\nut\nutb d\dbar$ & & & & & \\
41 & $\nut\nutb s\sbar$ & & & & & \\
42 & $\nut\nutb b\bbar$ & & & & & \\
\hline \hline
\end{tabular}
\end{center}
\caption[.]{semileptonic four-fermion final states in $\epl\emn$ collisions
  {\it (continued from previous page)}.}
\end{table}

\newpage
\begin{table}[h]
\begin{center}
\begin{tabular}{|l|l|c|c|c|c|l|} \hline\hline
label & final state & $N_a$ & $N_n$ & total & $N_b$ & signals \\
\hline
1 & \ru{$u\ubar d\dbar$} & 33 & 2 & 35 & 2 & 1 2 \\
2 & $c\cbar d\dbar$ & & & & &  \\
\hline
3 &\ru{$u\dbar s\cbar$} & 9 & 2 & 11 & 0 & 1 \\
4 & $d\ubar c\sbar$ & & & & & \\
\hline
5 & $u\ubar u\ubar$ & 48 & 0 & 48 & 20 & 2 \\
6 & $c\cbar c\cbar$ & & & & & \\
\hline
7 & \ru{$d\dbar d\dbar$} & 48 & 0 & 48 & 20 & 2 \\
8 & $s\sbar s\sbar$ & & & & & \\
9 & $b\bbar b\bbar$ & & & & & \\
\hline
10 & $u\ubar c\cbar$ & 24 & 0 & 24 & 10 & 2 \\
\hline
11 & $u\ubar s\sbar$ & 24 & 0 & 24 & 10 & 2 \\
12 & $u\ubar b\bbar$ & & & & & \\
13 & $c\cbar d\dbar$ & & & & & \\
14 & $c\cbar b\bbar$ & & & & & \\
\hline
15 & \ru{$d\dbar s\sbar$} & 24 & 0 & 24 & 10 & 2 \\
16 & $d\dbar b\bbar$ & & & & & \\
17 & $s\sbar b\bbar$ & & & & & \\
\hline \hline
\end{tabular}
\end{center}
\caption[.]{hadronic four-fermion final states in $\epl\emn$ collisions.}
\label{tablethree}
\end{table}

\newpage
\begin{table}[h]
\begin{center}
\begin{tabular}{|l|c|c|c|c|} \hline\hline
 &$\sqrt s= 165\,GeV$&$\sqrt s= 175\,GeV$&$\sqrt s= 185\,GeV$&
$\sqrt s= 195\,GeV$   \\ \hline
An.  &11.95         &18.48         &20.41         &20.82          \\ \hline
M. C.&11.96$\pm$0.03&18.45$\pm$0.04&20.40$\pm$0.05&20.79$\pm$0.05 \\
\hline \hline
 \end{tabular}
\end{center}
\caption[.]{analytic vs. Monte Carlo W-pair signal (in picobarns).\\
Here $\alpha= 1/129$, $M_W= 80.5~GeV$, $M_Z= 91.9~GeV$,
$\cos \theta_W= M_W/M_Z$ and $\Gamma_W= (3 \alpha M_W)/(4\sin^2 \theta_W)$.
The zero width extrapolated values of M.C. are obtained with the substitutions
$\Gamma_W \to \Gamma_W/N$ and $\sigma \to \sigma/N^2$, by taking
the numerical limit for growing N.}
\label{tablefour}
\end{table}

\newpage
\begin{table}[h]
\begin{center}
\begin{tabular}{|l|c|c|c|c|} \hline\hline
 process &$\sqs= 150\,GeV$ &  $\sqs= 175\,GeV$ & $\sqs= 200\,GeV$
  &$\sqs= 500\,GeV$ \\ \hline
W-pair       &\rrs{3600}{-2}&\err{1181}   &\err{1304}    &\rrs{2130}{-1}\\
             &\err{0011}    &\err{0002}   &\err{0003}    &\err{0011}  \\
\hline
$\epl\emn\nue\nueb$ &\rrs{2949}{-2}&\err{1208}    &\err{1466}
                    &\rrs{6111}{-1}\\
            &\err{0008}    &\err{0002}    &\err{0003}    &\err{0028}    \\
\hline
$\mpl\mmn\num\numb$ &\rrs{4350}{-2}&\err{1196}    &\err{1364}
                    &\rrs{2209}{-1}\\
            &\err{0012}    &\err{0002}    &\err{0003}    &\err{0011}    \\
\hline
$\epl\emn\num\numb$ &\rrs{8379}{-3}&\rrs{1906}{-2}&\rrs{6226}{-2}
                    &\rrs{2342}{-2}\\
            &\err{0026}    &\err{0004}    &\err{0008}    &\err{0006}    \\
\hline
$\nut\nutb\mpl\mmn$ &\rrs{7927}{-3}&\rrs{1342}{-2}&\rrs{5457}{-2}
                    &\rrs{1054}{-2}\\
            &\err{0031}    &\err{0003}    &\err{0007}    &\err{0003}    \\
\hline
$\nue\nueb\mpl\mmn$ &\rrs{1002}{-2}&\rrs{1842}{-2}&\rrs{6004}{-2}
                    &\rrs{7660}{-2}\\
            &\err{0004}    &\err{0005}    &\err{0009}    &\err{0079}    \\
\hline \hline
W-pair      &\rrs{3598}{-2}&\err{1187}   &\err{1315}    &\rrs{2130}{-1}\\
            &\err{0011}    &\err{0002}    &\err{0003}    &\err{0011}   \\
\hline
$\emn\nueb\num\mpl$ &\rrs{2912}{-2}&\err{1192}    &\err{1359}
                    &\rrs{3598}{-1}\\
            &\err{0008}    &\err{0002}    &\err{0003} &\err{0014}    \\
\hline
$\mmn\numb\nut\tpl$ &\rrs{3649}{-2}&\err{1193}    &\err{1307}
                    &\rrs{2101}{-1}\\
            &\err{0010}    &\err{0002}    &\err{0003} &\err{0010}    \\
\hline \hline
 \end{tabular}
\end{center}
\caption[.]{leptonic cross sections in picobarns. The second line of
each entry is the Monte Carlo error. In the first six entries
$m\p{\epl\emn},m\p{\mpl\mmn}>10~GeV$,
$|\cos \theta_{e^{\pm},~\mu^{\pm}}|<0.9$ ($\theta$ is the scattering
angle), $E_{e^{\pm},~\mu^{\pm}}>20~GeV$. In the last
three ones $E_{\emn,~\mu^{\pm},~\tpl}>20~GeV$,
$|\cos \theta_{\emn,~\mu^{\pm},~\tpl}|<0.9$.}
\label{tablefive}
\end{table}

\newpage
\begin{table}[h]
\begin{center}
\begin{tabular}{|l|c|c|c|c|} \hline\hline
process &$\sqs= 150\,GeV$ &  $\sqs= 175\,GeV$ & $\sqs= 200\,GeV$
        &$\sqs= 500\,GeV$ \\ \hline
W-pair  &\rrs{9605}{-2}&\err{3342} &\err{3601} &\rrs{4682}{-1}\\
                  &\err{0030}    &\err{0007}&\err{0008}&\err{0028}   \\
\hline
\ru{$\emn\nueb u\dbar$}&\rrs{7563}{-2}&\err{3345}&\err{3745}&\rrs{8293}{-1}\\
                  &\err{0021}    &\err{0007}&\err{0009}&\err{0035}    \\
\hline
\ru{$\mmn\numb u\dbar$}&\rrs{9728}{-2}&\err{3337}&\err{3610}&\rrs{4599}{-1}\\
                  &\err{0028}    &\err{0007}&\err{0008}&\err{0024}    \\
\hline \hline
 \end{tabular}
\end{center}
\caption[.]{semileptonic cross sections (in pb).
$E_{\emn,~\mmn,~u,~\dbar}>20~GeV$, $|\cos \theta_{\emn,~\mmn,~u,~\dbar}|<0.9$
, $|\cos \angle \p{u,\,\dbar}|<0.9$, $m\p{u\dbar}>10~GeV$.}
\label{tablesix}
\end{table}

\newpage
\begin{table}[h]
\begin{center}
\begin{tabular}{|l|c|c|c|} \hline\hline
   &$\sqs= 190\,GeV$ &  $\sqs= 200\,GeV$ & $\sqs= 500\,GeV$ \\
\hline
W-pair       &\err{3532}&\err{3419}&\rrs{2498}{-1}\\
             &\err{0009}&\err{0008}&\err{0018}    \\
1 cut        &\err{3612}&\err{3550}&\rrs{3522}{-1}\\
             &\err{0009}&\err{0009}&\err{0020}    \\
\hline
W-pair       &\err{3392}&\err{3259}&\rrs{2236}{-1}\\
             &\err{0009}&\err{0009}&\err{0014}    \\
2 cuts       &\err{3407}&\err{3292}&\rrs{2228}{-1}\\
             &\err{0009}&\err{0009}&\err{0014}    \\
\hline \hline
 \end{tabular}
\end{center}
\caption[.]{$\emn\p{3}\,\nueb\p{4}\,u\p{5}\,\dbar\p{6}$ cross section
in picobarns.
The $1^{st}$ entry is a comparison between the WW signal and the full result
with a cut $70<m_\p{56}<90~GeV$.
In the $2^{nd}$ entry $70<m_\p{34},m_\p{56}<90~GeV$.
All the other cuts are like in table \ref{tablesix},
 except that an energy cut $E_{\emn,~u,~\dbar}>50~GeV$
is required at $\sqs= 500\,GeV$.}
\label{tableseven}
\end{table}

\newpage
\begin{table}[h]
\begin{center}
\begin{tabular}{|l|c|c|c|c|} \hline\hline
process &$\sqs= 150\,GeV$ &  $\sqs= 175\,GeV$ & $\sqs= 200\,GeV$
        &$\sqs= 500\,GeV$ \\ \hline
W-pair&\rrs{2141}{-1}&\err{7699}   &\err{8726}    &\err{1083}  \\
             &\err{0019}    &\err{0018}   &\err{0023}    &\err{0008}  \\
\hline
\ru{$u\ubar d\dbar$}&\rrs{2317}{-1}&\err{7745} &\err{8987} &\err{1103}  \\
               &\err{0010}    &\err{0018}   &\err{0023}    &\err{0006}  \\
\hline
\ru{$u\dbar s\cbar$}&\rrs{2132}{-1}&\err{7720} &\err{8709} &\err{1052}  \\
               &\err{0010}    &\err{0018}   &\err{0023}    &\err{0006}  \\
\hline
$u\ubar u\ubar$&\rrs{1033}{-2}&\rrs{1956}{-2}&\rrs{1031}{-1}
                    &\rrs{1845}{-2}\\
               &\err{0008}    &\err{0006}    &\err{0002}    &\err{0006}  \\
\hline
\ru{$d\dbar d\dbar$}&\rrs{3900}{-3}&\rrs{1171}{-2}&\rrs{1508}{-1}
                    &\rrs{2239}{-2}\\
               &\err{0018}    &\err{0003}    &\err{0003}    &\err{0008}  \\
\hline
$u\ubar c\cbar$&\rrs{2417}{-2}&\rrs{4032}{-2}&\rrs{2066}{-1}&\rrs{3715}{-2}\\
               &\err{0012}    &\err{0014}    &\err{0004}    &\err{0013}  \\
\hline
$u\ubar s\sbar$&\rrs{1836}{-2}&\rrs{3474}{-2}&\rrs{2499}{-1}&\rrs{4143}{-2}\\
               &\err{0009}    &\err{0011}    &\err{0005}    &\err{0014}  \\
\hline
\ru{$d\dbar s\sbar$}&\rrs{8245}{-3}&\rrs{2437}{-2}&\rrs{3026}{-1}
                    &\rrs{4475}{-2}\\
               &\err{0038}    &\err{0007}    &\err{0006}    &\err{0017}  \\
\hline \hline
 \end{tabular}
\end{center}
\caption[.]{hadronic cross sections (in pb). $E_\p{all~particles}>20~GeV$,
$|\cos \theta_{\p{all~particles}}|<0.9$. Moreover $m\p{ij}>10~GeV$ and
$|\cos \angle \p{i,\,j}|<0.9$ between all possible final state couples.}
\label{tableeight}
\end{table}

\newpage
\begin{table}[h]
\begin{center}
\begin{tabular}{|c|c|c|c|c|} \hline\hline
$m_W\p{34}$&$m_W\p{56}$&$m_Z\p{36}$&$m_Z\p{45}$& cross section (pb)\\ \hline
0&0&0&0&\res{.6362}{.0017}{2} \\  \hline
0&0&0&1&\res{.1550}{.0007}{2} \\  \hline
0&0&1&0&\res{.5938}{.0024}{2} \\  \hline
0&0&1&1&\res{.4144}{.0038}{3} \\  \hline
0&1&0&0&\res{.1441}{.0002}{1} \\  \hline
0&1&0&1&\res{.7142}{.0045}{3} \\  \hline
0&1&1&0&\res{.8217}{.0075}{3} \\  \hline
0&1&1&1&\res{.2219}{.0089}{4} \\  \hline
1&0&0&0&\res{.1441}{.0002}{1} \\  \hline
1&0&0&1&\res{.7194}{.0045}{3} \\  \hline
1&0&1&0&\res{.8162}{.0074}{3} \\  \hline
1&0&1&1&\res{.2197}{.0086}{4} \\  \hline
1&1&0&0&\res{.1318}{.0003}{1} \\  \hline
1&1&0&1&\res{.6840}{.0066}{3} \\  \hline
1&1&1&0&\res{.1065}{.0011}{3} \\  \hline
1&1&1&1&      0  events       \\
\hline \hline
 \end{tabular}
\end{center}
\caption[.]{invariant mass analysis on $\emn\p{3}\,\nueb\p{4}\,
\nue\p{5}\,\epl\p{6}$ at $\sqs= 500\,GeV$.
In the first four columns a number 1 is written when the invariant mass
$m_V\p{ij}$ is inside the interval
$M_V-2\Gamma_V<m_V\p{ij}< M_V+2\Gamma_V$ (V= Z,W),
otherwise the corresponding entry is 0. Additional cuts are
$E_{\epl,~\emn}>20~GeV$,
$|\cos \theta_{\epl,~\emn}|<0.9$, $m\p{\epl\emn}>30~GeV$.}
\label{tablenine}
\end{table}

\newpage
\begin{table}[h]
\begin{center}
\begin{tabular}{|l|c|c|c|} \hline\hline
  &$\sqs= 500\,GeV$ &  $\sqs= 1000\,GeV$ & $\sqs= 2000\,GeV$ \\ \hline
$ \si  \p{1}   $&\res{.2127}{.0002}{1}&\res{.5885}{.0010}{2}
                &\res{.2045}{.0009}{2}\\
$ \sih \p{1}   $&\res{.1317}{.0002}{1}&\res{.3324}{.0005}{2}
                &\res{.8707}{.0022}{3}\\
$ \sith\p{1}   $&\res{.1318}{.0003}{1}&\res{.3339}{.0022}{2}
                &\res{.8917}{.0145}{3}\\
\hline
$ \si  \p{2}   $&\res{.6319}{.0004}{3}&\res{.1548}{.0002}{3}
                &\res{.4427}{.0010}{4}\\
$ \sih \p{2}   $&\res{.3921}{.0003}{3}&\res{.9426}{.0009}{4}
                &\res{.2589}{.0003}{4}\\
$ \sith\p{2}   $&\res{.4144}{.0038}{3}&\res{.1009}{.0015}{3}
                &\res{.2689}{.0077}{4}\\
\hline
$ \si  \p{3}   $&\res{.1169}{.0001}{1}&\res{.1201}{.0001}{1}
                &\res{.7250}{.0018}{2}\\
$ \sih \p{3}   $&\res{.8969}{.0007}{2}&\res{.9483}{.0009}{2}
                &\res{.5601}{.0007}{2}\\
$ \sith\p{3}   $&\res{.1441}{.0002}{1}&\res{.1161}{.0004}{1}
                &\res{.6126}{.0037}{2}\\
\hline
$ \si  \p{4}   $&\res{.1760}{.0001}{2}&\res{.1159}{.0001}{2}
                &\res{.6167}{.0004}{3}\\
$ \sih \p{4}   $&\res{.1379}{.0001}{2}&\res{.9449}{.0005}{3}
                &\res{.5103}{.0004}{3}\\
$ \sith\p{4}   $&\res{.1550}{.0007}{2}&\res{.1015}{.0009}{2}
                &\res{.5281}{.0060}{3}\\
\hline
$ \si  \p{5}   $&\res{.6840}{.0013}{2}&\res{.1732}{.0009}{1}
                &\res{.3038}{.0066}{1}\\
$ \sih \p{5}   $&\res{.5327}{.0012}{2}&\res{.1382}{.0007}{1}
                &\res{.2233}{.0031}{1}\\
$ \sith\p{5}   $&\res{.5938}{.0024}{2}&\res{.1411}{.0008}{1}
                &\res{.2275}{.0031}{1}\\
\hline
$\sum_i\si\p{i}$&\res{.4219}{.0004}{1}&\res{.3653}{.0010}{1}
                &\res{.4033}{.0069}{1}\\
$\sit          $&\res{.6017}{.0005}{1}&\res{.5099}{.0011}{1}
                &\res{.4459}{.0035}{1}\\
\hline \hline
 \end{tabular}
\end{center}
\caption[.]{cross sections for $\epl \emn \nue \nueb$ in picobarns.
$\si \p{i}$ are the cross sections for the processess (1)-(5)
(see introduction) computed including only the signal diagrams.
$\sih \p{i}$ are the
corresponding cross sections reconstructed using an invariant mass
analysis (see caption of table \ref{tablenine}). In $\sit$ all diagrams are
included and $\sith \p{i}$ are the results of the invariant mass analysis on
$\sit$. $E_{\epl,~\emn}>20~GeV$,
$|\cos \theta_{\epl,~\emn}|<0.9$, $m\p{\epl\emn}>30~GeV$.}
\label{tableten}
\end{table}

\newpage
\begin{table}[h]
\begin{center}
\begin{tabular}{|l|c|c|c|} \hline\hline
  &$\sqs= 500\,GeV$ &  $\sqs= 1000\,GeV$ & $\sqs= 2000\,GeV$ \\ \hline
$ \si  \p{1}   $&\res{.4702}{.0006}{1}&\res{.1428}{.0003}{1}
                &\res{.5460}{.0025}{2}\\
$ \sih \p{1}   $&\res{.2823}{.0004}{1}&\res{.7907}{.0014}{2}
                &\res{.2272}{.0006}{2}\\
$ \sith\p{1}   $&\res{.2822}{.0005}{1}&\res{.7920}{.0020}{2}
                &\res{.2275}{.0003}{2}\\
\hline
$ \si  \p{3}   $&\res{.2961}{.0002}{1}&\res{.2953}{.0003}{1}
                &\res{.1817}{.0001}{1}\\
$ \sih \p{3}   $&\res{.2279}{.0002}{1}&\res{.2348}{.0002}{1}
                &\res{.1417}{.0001}{1}\\
$ \sith\p{3}   $&\res{.3516}{.0004}{1}&\res{.2836}{.0003}{1}
                &\res{.1559}{.0001}{1}\\
\hline
$\sum_i\si\p{i}$&\res{.7663}{.0008}{1}&\res{.4381}{.0006}{1}
                &\res{.2363}{.0004}{1}\\
$\sit          $&\res{.8273}{.0007}{1}&\res{.4520}{.0004}{1}
                &\res{.2192}{.0001}{1}\\
\hline \hline
 \end{tabular}
\end{center}
\caption[.]{cross sections for $\emn \nueb u \dbar$ (pb). See caption of
table \ref{tableten}. $E_{\emn,~u,~\dbar}>20~GeV$,
$|\cos \theta_{\emn,~u,~\dbar}|<0.9$, $m\p{u\dbar}>30~GeV$.}
\label{tableeleven}
\end{table}

\newpage
\begin{table}[h]
\begin{center}
\begin{tabular}{|l|c|c|c|} \hline\hline
  &$\sqs= 500\,GeV$ &  $\sqs= 1000\,GeV$ & $\sqs= 2000\,GeV$ \\ \hline
$ \si  \p{2}   $&\res{.7298}{.0002}{3}&\res{.2074}{.0001}{3}
                &\res{.6386}{.0004}{4}\\
$ \sih \p{2}   $&\res{.4536}{.0001}{3}&\res{.1301}{.0001}{3}
                &\res{.3930}{.0002}{4}\\
$ \sith\p{2}   $&\res{.4765}{.0008}{3}&\res{.1351}{.0003}{3}
                &\res{.4015}{.0012}{4}\\
\hline
$ \si  \p{4}   $&\res{.1725}{.0001}{2}&\res{.9679}{.0002}{3}
                &\res{.4775}{.0001}{3}\\
$ \sih \p{4}   $&\res{.1354}{.0001}{2}&\res{.7874}{.0002}{3}
                &\res{.3924}{.0001}{3}\\
$ \sith\p{4}   $&\res{.1361}{.0001}{2}&\res{.7895}{.0006}{3}
                &\res{.3919}{.0003}{3}\\
\hline
$\sum_i\si\p{i}$&\res{.2455}{.0001}{2}&\res{.1175}{.0001}{2}
                &\res{.5414}{.0001}{3}\\
$\sit          $&\res{.2833}{.0001}{2}&\res{.1398}{.0001}{2}
                &\res{.6471}{.0004}{3}\\
\hline \hline
 \end{tabular}
\end{center}
\caption[.]{cross sections for $\epl \emn  u \ubar$ in picobarns.
See caption of table \ref{tableten}.  $E_\p{all~particles}>20~GeV$,
$|\cos \theta_{\p{all~particles}}|<0.9$, $~m\p{u\ubar},m\p{\epl\emn}>30~GeV$.}
\label{tabletwelve}
\end{table}

\newpage
\begin{table}[h]
\begin{center}
\begin{tabular}{|l|l|l|l|} \hline\hline
 &$~~~\sqs= 500\,GeV$ & $~~~\sqs= 1000\,GeV$ & $~~~\sqs= 2000\,GeV$ \\ \hline
$ \si  \p{2}   $&\res{.2165}{.0001}{2}&\res{.5306}{.0006}{3}
                &\res{.1507}{.0003}{3}\\
$ \sih \p{2}   $&\res{.1334}{.0001}{2}&\res{.3219}{.0003}{3}
                &\res{.8848}{.0013}{4}\\
$ \sith\p{2}   $&\res{.1364}{.0006}{2}&\res{.3273}{.0045}{3}
                &\res{.9275}{.0282}{4}\\
\hline
$ \si  \p{5}   $&\res{.2242}{.0005}{1}&\res{.5795}{.0029}{1}
                &$.1331 \pm .0324$\\
$ \sih \p{5}   $&\res{.1749}{.0004}{1}&\res{.4601}{.0025}{1}
                &\res{.7631}{.0101}{1}\\
$ \sith\p{5}   $&\res{.1680}{.0004}{1}&\res{.4554}{.0025}{1}
                &\res{.7781}{.0104}{1}\\
\hline
$\sum_i\si\p{i}$&\res{.2459}{.0005}{1}&\res{.5848}{.0029}{1}
                &$.1333 \pm .0324$\\
$\sit          $&\res{.2374}{.0004}{1}&\res{.5663}{.0027}{1}
                &\res{.9301}{.0110}{1}\\
\hline \hline
 \end{tabular}
\end{center}
\caption[.]{Cross sections for $\nue \nueb  u \ubar$ (pb). See caption of
table \ref{tableten}. $E_{u,~\ubar}>20~GeV$, $|\cos \theta_{u,~\ubar}|<0.9$,
$m\p{u\ubar}>30~GeV$.}
\label{tablethirteen}
\end{table}
\end{document}